\documentclass[a4paper,11pt]{article}

\usepackage{subfig}
\usepackage{diagbox}
\usepackage{jinstpub}

\title{Horizontal Position Reconstruction in PandaX-II}
\collaboration{PandaX Collaboration}
\author[b,1]{Dan Zhang, \note{corresponding author}}
\author[b,2]{Andi Tan, \note{corresponding author, now at Department of Physics, Princeton University, Princeton, NJ, USA}}
\author[a]{Abdusalam Abdukerim,}
\author[a]{Wei Chen,}
\author[a,c]{Xun Chen,}
\author[d]{Yunhua Chen,}
\author[e]{Chen Cheng,}
\author[f]{Xiangyi Cui,}
\author[g]{Yingjie Fan,}
\author[h]{Deqing Fang,}
\author[h]{Changbo Fu,}
\author[i]{Mengting Fu,}
\author[j,k]{Lisheng Geng,}
\author[a]{Karl Giboni,}
\author[a]{Linhui Gu,}
\author[d]{Xuyuan Guo,}
\author[a]{Ke Han,}
\author[a]{Changda He,}
\author[d]{Shengming He,}
\author[a]{Di Huang,}
\author[d]{Yan Huang,}
\author[l]{Yanlin Huang,}
\author[a]{Zhou Huang,}
\author[b]{Xiangdong Ji,}
\author[m]{Yonglin Ju,}
\author[f]{Shuaijie Li,}
\author[n,o]{Qing Lin,}
\author[m]{Huaxuan Liu,}
\author[a,c,f,3]{Jianglai Liu, \note{spokesperson, jianglai.liu@sjtu.edu.cn}}
\author[d]{Liqiang Liu,}
\author[p,q]{Xiaoying Lu,}
\author[a]{Wenbo Ma,}
\author[c,h]{Yugang Ma,}
\author[i]{Yajun Mao,}
\author[a,d]{Yue Meng,}
\author[a]{Parinya Namwongsa,}
\author[a]{Kaixiang Ni,}
\author[d]{Jinhua Ning,}
\author[a]{Xuyang Ning,}
\author[p,q]{Xiangxiang Ren,}
\author[p,q]{Nasir Shaheed,}
\author[d]{Changsong Shang,}
\author[j]{Guofang Shen,}
\author[a]{Lin Si,}
\author[p,q]{Anqing Wang,}
\author[r]{Hongwei Wang,}
\author[p,q]{Meng Wang,}
\author[h]{Qiuhong Wang,}
\author[i]{Siguang Wang,}
\author[e]{Wei Wang,}
\author[m]{Xiuli Wang,}
\author[a,c]{Zhou Wang,}
\author[e]{Mengmeng Wu,}
\author[d]{Shiyong Wu,}
\author[a]{Weihao Wu,}
\author[a]{Jingkai Xia,}
\author[b]{Mengjiao Xiao,}
\author[e]{Xiang Xiao,}
\author[f]{Pengwei Xie,}
\author[a,c]{Binbin Yan,}
\author[a]{Jijun Yang,}
\author[a]{Yong Yang,}
\author[g]{Chunxu Yu,}
\author[p,q]{Jumin Yuan,}
\author[a]{Ying Yuan,}
\author[a]{Xinning Zeng,}
\author[a,c]{Tao Zhang,}
\author[a,c]{Li Zhao,}
\author[l]{Qibin Zheng,}
\author[d]{Jifang Zhou,}
\author[a]{Ning Zhou,}
\author[j]{Xiaopeng Zhou}

\affiliation[a]{School of Physics and Astronomy, Shanghai Jiao Tong University, MOE Key Laboratory for Particle Astrophysics and Cosmology, Shanghai Key Laboratory for Particle Physics and Cosmology, Shanghai 200240, China}
\affiliation[b]{Department of Physics, University of Maryland, College Park, Maryland 20742, USA}
\affiliation[c]{Shanghai Jiao Tong University Sichuan Research Institute, Chengdu 610213, China}
\affiliation[d]{Yalong River Hydropower Development Company, Ltd., 288 Shuanglin Road, Chengdu 610051, China}

\affiliation[e]{School of Physics, Sun Yat-Sen University, Guangzhou 510275, China }
\affiliation[f]{Tsung-Dao Lee Institute, Shanghai 200240, China}
\affiliation[g]{School of Physics, Nankai University, Tianjin 300071, China}
\affiliation[h]{Key Laboratory of Nuclear Physics and Ion-beam Application (MOE), Institute of Modern Physics, Fudan University, Shanghai 200433, China}

\affiliation[i]{School of Physics, Peking University, Beijing 100871, China}
\affiliation[j]{School of Physics, Beihang University, Beijing 100191, China}
\affiliation[k]{International Research Center for Nuclei and Particles in the Cosmos \& Beijing Key Laboratory of Advanced Nuclear Materials and Physics, Beihang University, Beijing 100191, China}
\affiliation[l]{School of Medical Instrument and Food Engineering, University of Shanghai for Science and Technology, Shanghai 200093, China}
\affiliation[m]{School of Mechanical Engineering, Shanghai Jiao Tong University, Shanghai 200240, China}
\affiliation[n]{State Key Laboratory of Particle Detection and Electronics,
University of Science and Technology of China, Hefei 230026, China}
\affiliation[o]{Department of Modern Physics, University of Science and Technology of China, Hefei 230026, China}
\affiliation[p]{Key Laboratory of Particle Physics and Particle Irradiation (MOE), Shandong University, Jinan 250100, China}
\affiliation[q]{Research Center for Particle Science and Technology, Institute of Frontier and Interdisciplinary Science, Shandong University, Qingdao 266237, Shandong, China}
\affiliation[r]{Shanghai Advanced Research Institute, Chinese Academy of Sciences, Shanghai 201210, China}

\emailAdd{dzhang16@umd.edu, andytan@umd.edu}
\abstract{Dual-phase noble-gas time projection chambers (TPCs) have improved the sensitivities for dark matter direct search in past decades. The capability of TPCs to reconstruct 3-D vertexes of keV scale recoilings is one of the most advantageous features.
In this work, we develop two horizontal position reconstruction algorithms for the PandaX-II dark matter search experiment using the dual-phase liquid xenon TPC. Both algorithms are optimized by the $^{83m}$Kr calibration events and use photon distribution of ionization signals among photomultiplier tubes to infer the positions. According to the events coming from the gate electrode, the uncertainties in the horizontal positions are 3.4~mm (3.9~mm) in the analytical (simulation-based) algorithm for an ionization signal with several thousand photon electrons in the center of the TPC.}

\keywords{Time Projection Chamber, Photomultiplier, Horizontal Position Reconstruction.}

\begin{document}
\maketitle
\flushbottom

\section{Introduction}
%
%
%
%
Weakly interacting massive particles (WIMPs)  are a class of hypothetical particles to explain the nature of dark matter (DM) in the astrophysical and cosmological observations~\cite{robin,bullet,planck}. 
The noble-gas detector has been one of the most sensitive WIMP-nucleus  scattering search methods for over a decade because of the scalability and strong background suppression~\cite{xenon10,lux2017,main2,recentMain,xenon1t,lz,darkside,dokeElena,bbyERmodel,karl}. 

 
Typically, a dual-phase noble-gas time projection chamber~(TPC) has photon sensors on the top and at the bottom, and a vertical drifting electric field ($E_{\rm drift}$) in the liquid phase (Fig.~\ref{fig:tpc}). A recoiling event produces prompt scintillation photons ($S1$) and free electrons in the liquid. The electrons are drifted upward to the liquid-gas interface, where a stronger electric field ($E_{\rm extraction}$) extracts the electrons into the gas and generates proportional scintillation ($S2$). 

The spatial information of the events plays an important role in understanding the recoiling events and suppressing backgrounds.
For instance, with the scattering angles of the mono-energetic incoming neutrons known, nuclear recoil energy calibration is pushed down to 1~keV in the LUX experiment~\cite{ddlux}. 
More importantly, the spatial information suppresses the gamma and neutron backgrounds coming from outside of the sensitive region because the shielding effect of noble liquids leads to a strong spatial dependence in these backgrounds~\cite{neutron,qiuhongNeutron}.
Similarly, the surface backgrounds due to radioactivities attached to the materials can also be suppressed with positions known~\cite{neutron}.
Therefore, a more accurate position reconstruction brings potentials to improve the sensitivity of dark matter searching.


This work focuses on the horizontal position reconstruction in the PandaX-II detector where Hamamatsu-R11410 3-inch photomultiplier tubes (PMTs) are used as light sensors.
The vertical vertexes, reconstructed using the delay time of $S2$, can reach a resolution of a few millimeters with a drifting field $\sim \mathcal{O}(100~{\rm V/cm})$~\cite{dokeElena}. 
To reach a horizontal position resolution comparable to the vertical one, simple reconstruction using the center position of the hottest PMT is not sufficient because the PMTs are at least 8~cm apart. Sophisticated algorithms based on the distribution of the $S2$ collected by the PMTs are applied.

We develop two algorithms in this work based on the photon acceptance functions~(PAFs), $\eta$, which describe the light fraction collected by light sensors for one event as introduced in the literature~\cite{PAF1976}.
PAF is a function of light sensor index, $i$, and position of events $(x,y,z)$, because of the change in the solid angle subtended by the $i$th light sensor to $(x,y,z)$. It can be evaluated as
\begin{equation}
    \eta_{i}(x,y,z)= {\frac{{\rm~photons~detected~by~sensor}_{i}{\rm~for~an~event~happens~at~}(x,y,z)}{{\rm~total~photons~detected~for~the~event~happens~at~}(x,y,z)}}.
\end{equation}
During modeling, the functions are usually built with analytical models~\cite{zeplin1,posLux} or Monte Carlo simulations~\cite{template,posDS,xenon3d}, and trained with calibration data. 
Generally, a better agreement between the model and reconstructed data PAFs leads to a better position reconstruction.

{\color{black}To optimize the models, calibration events with positions or distributions known and energies close to WIMP searches are favored. In the PandaX-II liquid xenon detector, we use $^{83m}$Kr isotopes released by the customized $^{83}$Rb ($T_{1/2}=86.2$~d) sources and flushed into our TPC~\cite{kr83m}. The $^{83m}$Kr isotopes are uniformly distributed in the detector because of a long enough lifetime ($T_{1/2}=1.83$~h) to mix with xenon. In addition, the $^{83m}$Kr events provide signals close to a typical WIMP search window $\sim \mathcal{O}(10~{\rm keV})$. $^{83m}$Kr isotopes released by $^{83}$Rb decay into the ground state with two successive transitions of 32.1~keV and 9.4~keV. Because separating the two with a 154~ns intervening half-life is difficult, we use the sum of the transitions.}

Our developments on analytical and simulation-based PAFs are both tuned with $^{83m}$Kr calibration data.  
The detailed geometry of the PandaX-II detector is described in \cite{prd}. 
The sensitive volume is surrounded by a dodecagonal polytetrafluoroethylene~(PTFE) surface, and covered with 55 PMTs at the bottom and on the top, respectively, with the same alignment.
In the analytical algorithm, we extend the single-variable PAF based on ZEPLIN III's work to correct the light reflection effect because of the PTFE surface~\cite{zeplin1}. 
In the simulation-based algorithm, We tune the vertical vertexes of $S2s$ in the gas phase as a function of the horizontal position to make the simulation agree with calibration data.

We reach comparable quality in the two algorithms for the WIMP search purpose with $S2\in (100,10000)$~photo electron (PE).
With respect to uncertainties, the analytical algorithm is better for $S2s$ with several thousand PEs in the PandaX-II TPC, but becomes slightly worse than the simulation-based one for S2s with several hundred PEs as studied by the surface events from the PTFE panels surrounding the sensitive region.
The uncertainties in the center region of the TPC are $3.4$ ($3.9$)~mm in the analytical (simulation-based) algorithm as estimated by the radioactivities on the gate grid wires with $S2s$ larger than 1000~PE.
Apart from the uncertainties, the analytical algorithm presents slightly better uniformity, which is evaluated by the radial distribution of $^{83m}$Kr events. But for robustness, the simulation-based algorithm is more stable  when handling inhibited PMTs.


In this paper, the content is organized as follows. In Sec.~\ref{sec:paf}, we demonstrate the procedure to reconstruct positions with PAFs.
In Sec.~\ref{sec:ana} and Sec.~\ref{sec:simu}, we present the setup of refined analytical and simulation-based PAF with $^{83m}$Kr events in the PandaX-II detector sequentially. Finally, in Sec.~\ref{sec:comp}, we compare the two algorithms in the gate events and surface events besides $^{83m}$Kr events.

\section{Reconstruction procedure with PAF} \label{sec:paf}

\begin{figure}[!htb]
\centering
\includegraphics[width = 3.3in]{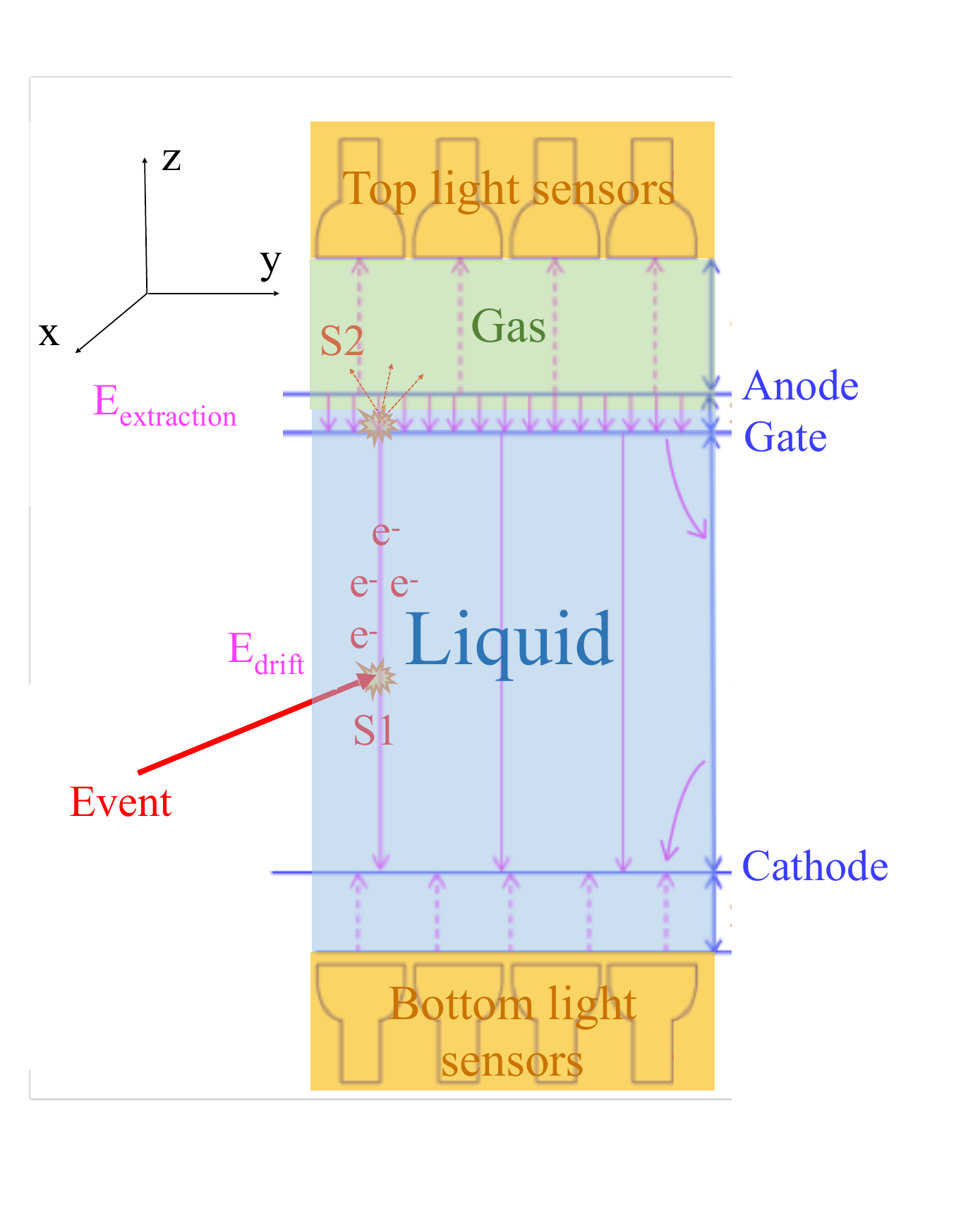}
\caption{\color{black} Sketch of a dual-phase nobel gas TPC. The two arrays of light sensors collect both prompt $S1s$ and delayed $S2s$. $E_{\rm drift}$ and $E_{\rm extraction}$ with their directions indicated by the magenta arrows are established with the electric potential differences among the anode, cathode and gate electrode. More complicated designs for a real TPC are not included in this sketch. The Cartesian coordinate marked on the upper-left corner is used throughout this work where $z$ is for vertical vertexes and $(x,y)$ for horizontal ones.}
\label{fig:tpc}
\end{figure}

PAF$_{i}$, describing the light fraction detected by the $i$th PMT in one event, is a function of the 3D scintillation position. In a TPC as Fig.~\ref{fig:tpc}, the functions depend on two-dimensional~(2D) horizontal positions because $S2s$ are all generated at the liquid-gas interface. We note the PAF$_{i}$ as $\eta_{i}(x,y)$ where $x$ and $y$ represent the horizontal vertexes, where $i$ only includes the top PMT array close to the interface.

Modeling of the PAFs is the first step. In this paper, we define two groups of PAFs, which are the model, $\eta_{i,\rm{model~=ana~or~simu}}(x,y)$, and data, $\eta_{i,\rm{data}}(x_{\rm rec},y_{\rm rec})$, PAFs. Analyitcal and simulation-based algorithms both model PAFs
with some adjustable parameters. The data PAFs can only be calculated with $x_{\rm rec}$ and $y_{\rm rec}$ reconstructed. The parameters are optimized for a better agreement between the model and data PAFs.

After the model building, we construct the likelihood function which is maximized by scanning possible $x$ and $y$ to infer the position of an event.
The likelihood function should reflect how charges statistically distribute among the PMTs. More specifically, in each PMT, the photons collected follow a Poisson distribution. The total likelihood function is a multiplication of a series of the Poisson distributions as a function of $x$ and $y$. 

The statistical inference of the position with input photons of an event is done by maximum likelihood (ML) estimation. 
As deduced in the literature~\cite{PAF1976,zeplin1}, maximizing the total likelihood function is equivalent to maximizing the simplified log likelihood,
\begin{equation}\label{equ:ml}
\ln{L}(x,y)=\sum_{i}S2_{i}\cdot\ln\frac{\eta_{i,{\rm{model~=~ana~or~simu}}}(x,y)}{P(x,y)},
\end{equation}
where  $P(x,y)=\sum_{i}{\eta_{i,\rm{model~=~ana~or~ simu}}(x,y)}$, and the summation includes all the top PMTs turned on.
The $\ln{L}$ is maximized by scanning $x$ and $y$ in every event with the $S2$ charges collected by PMT$_{i}$, $\{S2_{i}\}$, as inputs.

\section{Refined analytical PAF}\label{sec:ana}

To model the PAFs analytically,
PAFs were simplified as a single-variable function of the distance to the center of the PMT, $\iota$, in \cite{medical,zeplin1}. Qualitatively, a PAF is a monotonically decreasing function of $\iota$. PAFs are Gaussian distributions with the first-order corrections in \cite{medical}. ZEPLIN-III's work combines Cauchy and Gaussian distributions in modeling PAFs for a better agreement~\cite{zeplin1}. Later, LUX proposed a more sophisticated analytical model with $x$ and $y$ as variables based on the simulation of detector geometry and accounts asymmetries of the PMT positions according to the layout of PMTs and the horizontal boundary shape~\cite{posLux}. Because of a different detector geometry and PMT layout, we cannot use their modeled functions directly. Instead, we develop an extended single-variable model based on ZEPLIN-III's work~\cite{zeplin1}.

\subsection{Extended single-variable PAF}
The single-variable PAF used in ZEPLIN-III is kept as the basic analytic form~\cite{zeplin1},
\begin{equation}\label{equ:ana}
\eta_{i,\rm{ana}}^{0}(\iota_{i})=A_{i}\cdot \exp\left({-\frac{a_{i}\cdot\frac{\iota_{i}}{r_{i}}}{1+{(\frac{\iota_{i}}{r_{i}})}^{1-\alpha_{i}}}-\frac{b_{i}}{1+{(\frac{\iota_{i}}{r_{i}})}^{-\alpha_{i}}}}\right),
\end{equation}
where $i$ indicates the PMT index. The parameters, including $A_{i}$, $\alpha_{i}$, $r_{i}$, $a_{i}$ and $b_{i}$, are fitting parameters. The $\iota_{i}$ is the distance of the scattering point to the center of the PMT$_{i}$, 
\begin{equation}\label{equ:iota}
\iota_{i}(x,y) = \sqrt{(x-X_{i})^2 + (y-Y_{i})^2},
\end{equation}
where $(X_{i},Y_{i})$ is the center of PMT$_{i}$.

\begin{figure}[!htb]
\centering
\includegraphics[width = 3.5in]{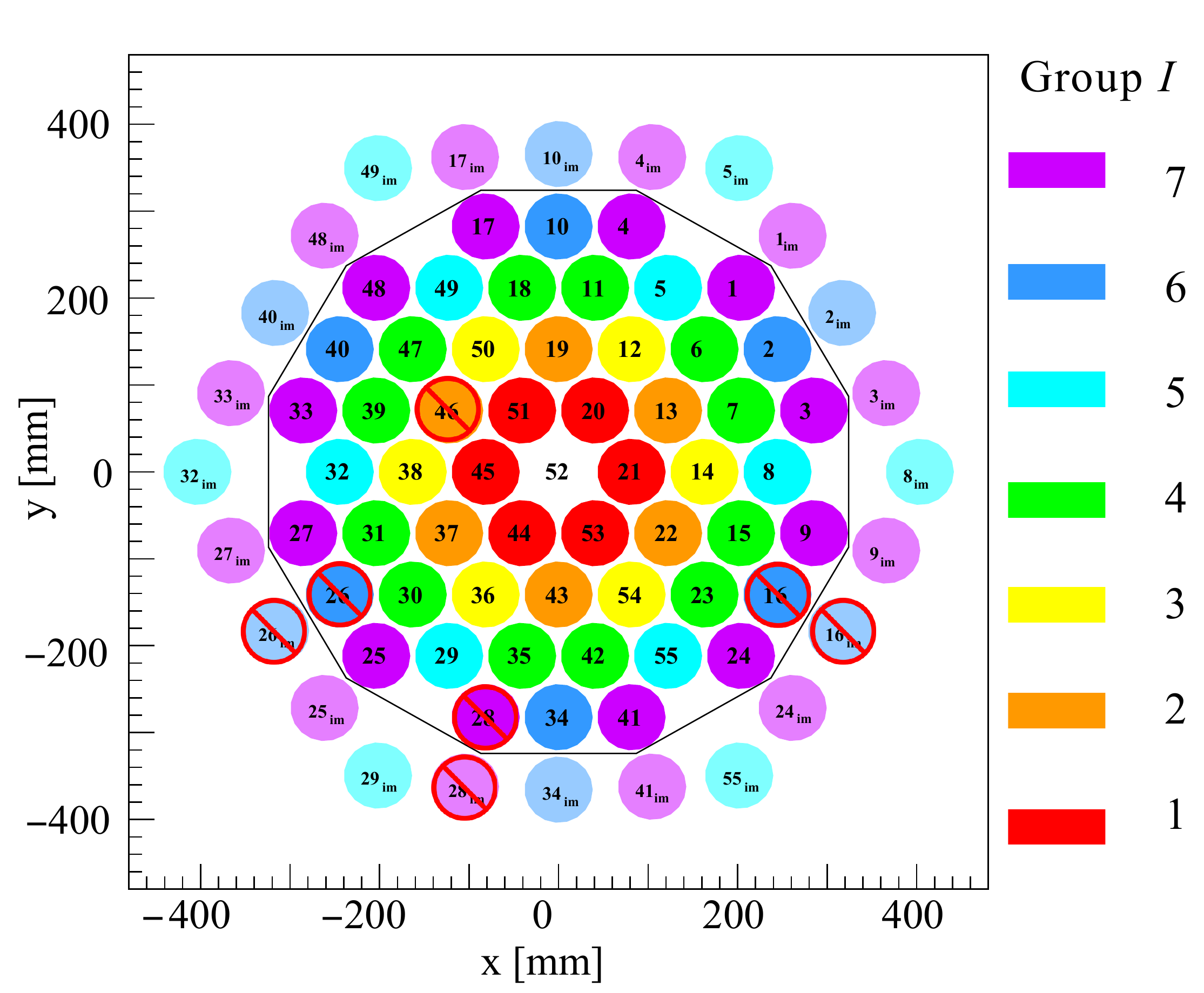}
\caption{The arrangement of the top PMTs in the PandaX-II detector. The distance between two closest PMTs is 81.5~mm. The average diameter of the detector is 658~mm. The ones marked as banned are not included in the position reconstruction, those with semi-transparent color are image PMTs to account for the reflections on the PTFE field cage wall surfaces.}
\label{fig:pmtArray}
\end{figure}

ZEPLIN's model which uses Eq.~\ref{equ:ana} as $\eta_{i,\rm{model}}$ causes problem at the large radius area for the PandaX-II detector. Without further adjustment, the model causes the events close to the PTFE surface to be congregated at the outermost PMT center which is around 3~cm away from the authentic surface. This inward bias is similar to the center-of-gravity algorithm due to non-optimized weights for $S2s$ collected by different PMTs.

To fix this problem and correct the asymmetry brought by the PTFE reflection at the border, we introduce the image PMTs using the similar concept of `image charge', which change the weights in the ML estimation (Fig.~\ref{fig:pmtArray}). The $\eta_{i,\rm{ana}}^{0}(\iota_{i})$ is separated into two parts, the reduced object, $\eta_{i,\rm{ana}}(\iota_{i})$, and the corresponding image, $\eta_{i,\rm{ana, im}}(\iota_{i,im})$. 
Mathematically, we add two groups of parameters, $\{w_{I}\}$ and $\{\rho_{I}\}$,
\begin{equation}\label{equ:extend}
\begin{split}
   & \eta_{i,\rm{ana}}(\iota_{i}) = \frac{1}{1+w_{I}} \cdot (1-\rho_{I})\cdot \eta_{i,\rm{ana}}^{0}(\iota_{i}),~{\rm and}\\
 &   \eta_{i,\rm{ana, im}}(\iota_{i,im}) = w_{I}\cdot\eta_{i,\rm{ana}}(\iota_{i,im}),
\end{split}
\end{equation}
where $I$ indicates the group number of the PMT, which is determined by the distance to the center of the TPC as in Fig.~\ref{fig:pmtArray}. 
The factor before $\eta_{i,\rm{ana}}^{0}(\iota_{i})$ in Eq.~\ref{equ:extend} suggests that the light collected by the edge PMT is shared with the image PMT. 
The $\{w_{I}\}$ is only nontrivial for the PMTs next to the boundary and zero for the inner PMTs. 
We reduce  $\{w_{I}\}$ to one parameter $w_{e}$, which is the same for all the three outmost groups ($I=5,6,7$) in the PandaX-II detector. 
The other parameter group, $\{\rho_{I}\}$, is nontrivial for each $I$ which helps to correct the global reflection effect. The minus sign before the $\rho_{I}$ in Eq.~\ref{equ:extend} represents a cut-off correction.

To include the image PMTs in the likelihood function, we modify Eq.~\ref{equ:ml} as follows,
\begin{equation}\label{equ:mlana}
\begin{aligned}
\ln{L}(x,y)= &
\sum_{\rm{edge}}\left(\frac{1}{1+w_{e}}\cdot S2_{i}\cdot\ln\frac{\eta_{i,\rm{ana}}(\iota_{i})}{P(x,y)} \right.\\
&+ \left.
 \frac{w_{e}}{1+w_{e}}\cdot S2_{i}\cdot\ln\frac{\eta_{i,\rm{ana,im}}(\iota_{i,\rm{im}})}{P(x,y)}\right)\\
&+\sum_{\rm{inner}}S2_{i}\cdot\ln\frac{\eta_{i,\rm{ana}}(\iota_{i})}{P(x,y)},
\end{aligned}
\end{equation}
where $P(x,y)=\sum_{i}\left({\eta_{i,\rm{ana}}}+\eta_{i,\rm{im,ana}}\right)$.


\subsection{Model training with $^{83m}$Kr in PandaX-II}
We scan the group parameters, $\{ w_{e},\rho_{I}\}$, to optimize the analytical model. The fitting parameters of all the PAFs, $\{A_{i}$, $\alpha_{i}$, $r_{i}$, $a_{i}$, $b_{i}\}$,  are initially set the same for each PMT$_i$ and then updated with iterative fittings. The quality of each group parameter is evaluated after reaching stable fitting results.


We parametrize $\{\rho_{I}\}$ for more efficient training.
The required computational resources increase exponentially with the number of the groups, $N_{g}$, because we have to scan parameters in $N_{g} + 1$ dimensions. 
Therefore, we further parametrize $\rho_{I}$ according to the group number and reduce the number of parameters to two in $\{\rho_{I}\}$ by requiring
 \begin{equation}\label{equ:rho}
\rho_{I}=\left\{
\begin{array}{l}
c\cdot({\frac{R_{i}}{R_{\rm{max}}}})^{4}, I=1,2,3\\
d\cdot({\frac{R_{i}}{R_{\rm{max}}}})^{4}, I=4,5,6,7
\end{array},
\right. R_{i}=\sqrt{X_{i}^2+Y_{i}^2},
\end{equation} 
where $R_{i}$ is the distance of the PMT$_{i}$ center to the TPC center, and $R_{\rm{max}}$ ($294$~mm) is $R_{7}$. In fact, the set $\{\rho_{I}\}$ is a higher order correction of the $\eta_{i,\rm{ana}}^{0}$, which can be expanded as Taylor series. {\color{black}We have tried different positive power numbers in the modeling. The uniformity of $^{83m}$Kr is used to evaluate the quality similar to the parameter optimization procedure discussed later in this section. The $R_{i}^4$ dependence is selected for application in the PandaX-II detector. }.

\begin{figure}[!htb]
\centering
\includegraphics[width = 3.3in]{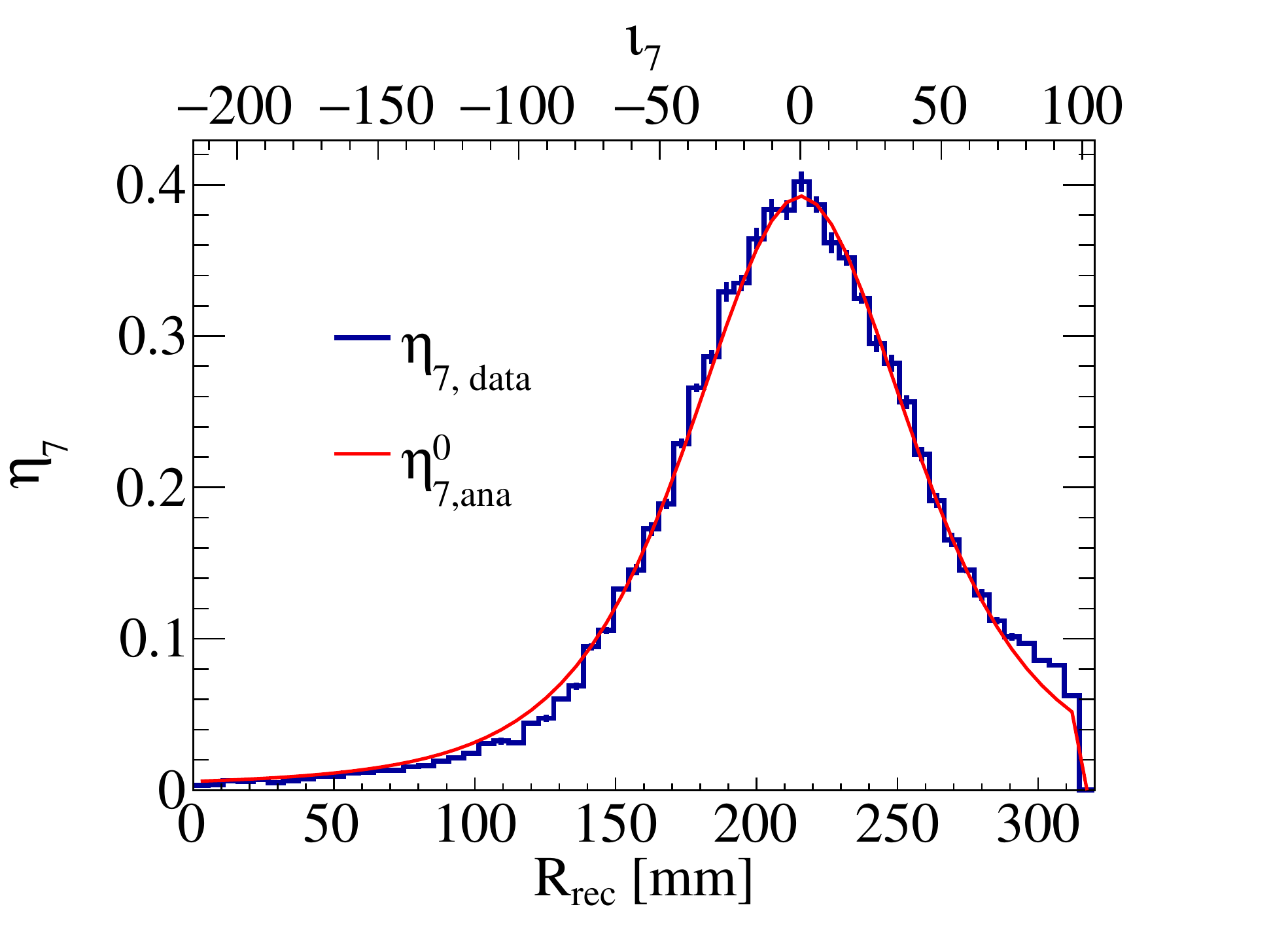}
\caption{An example of the extended PAF in PMT$_{7}$ in Fig.~\ref{fig:pmtArray}. The blue line is the $\eta_{7,\rm{data}}$ along the line from the origin to the center of PMT$_{7}$, $(X_7,Y_7)=(204,71)$~mm, and the red line is the fitted $\eta_{7,\rm{ana}}^{0}$.}
\label{fig:anaFitEx}
\end{figure}

We tune the outer and inner parameters sequentially based on the uniformity of $^{83m}$Kr. 
We scan $d$ and $w_{e}$ first as they have a larger effect on the uniformity. 
The initial fitting parameters of all the PAFs are set as the same. An example is $A_{i}=0.4$, $a_i=-0.47$, $b_i=6$, $\alpha_{i}=2.3$, $r_{i}=95$~mm. {\color{black} $A_{i}$ is the maximum of the PAF$_{i}$ at $\iota_{i} = 0$, $r_{i}$ reflects the size of the PMT and the other parameters are more phenomenological. Different initial values can be used as long as the fittings converge.}
With set group parameters and initial fitting parameters, the initial $\eta_{i,\rm{ana}}$ are set, and we can reconstruct positions by ML in Eq.~\ref{equ:mlana}. 
Instead of initializing coordinates with arbitrary numbers, we take the positions reconstructed by the center-of-gravity algorithm as initial positions, which makes the ML estimation faster and avoids taking local maxima for most events. 
After the first reconstruction, the new coordinates of $^{83m}$Kr are used to generate a data PAF as
\begin{equation}\label{equ:data}
{\eta}_{i,\rm{data}}(x_{\rm rec},y_{\rm rec})= \overline{ \frac{S2_{i}}{S2_{\rm{top}}}}(x_{\rm rec},y_{\rm rec}),
\end{equation}
where $S2_{\rm{top}}=\sum_{i}S2_{i}$  and the summation only includes the top PMTs. As 4 out of 55 PMTs are turned off due to sparkings and afterpulses (Fig.~\ref{fig:pmtArray}), 51 PMTs are in the summation.
The overline means averaging over the events in the same $x_{\rm rec}$-$y_{\rm rec}$ bin. 
Only ${\eta}_{i,\rm{data}}(x_{\rm rec},y_{\rm rec})$ along the line from the center of the TPC to $(X_{i},Y_{i})$ is used to fit the next $\eta^{0}_{i,\rm{ana}}(\iota_i)$ in Eq.~\ref{equ:ana}.
An example of the extended PAF fitting for PMT$_{7}$ is shown in  Fig.~\ref{fig:anaFitEx}, where $\eta_{7,\rm{ana}}^{0}(\iota_{7})$ in Eq.~\ref{equ:ana} is fitted to the $^{83m}$Kr data with $(x_{\rm rec},y_{\rm rec})$. $R_{\rm rec}$ (=$\sqrt{x_{\rm rec}^2+y_{\rm rec}^2}$) in Fig.~\ref{fig:anaFitEx} goes through the TPC center $(0,0)$ and the PMT$_{7}$ center $(X_7,Y_7)=(204,71)$~mm.
The five fitting parameters for each PMT$_{i}$ are updated in the new fitting and used to generate new positions with Eq.~\ref{equ:mlana}. 
Six iterations can reach consistent fitting results within a 1~mm difference, which takes several hours in total for one $(c,d,w_{e})$.

Following this, we estimate the quality of the reconstruction with the spatial distribution of the $^{83m}$Kr events. Another intuitive choice is the PTFE surface events like $^{210}$Po which may reflect the main problem. However, there is a risk of pushing the surface events to the PTFE position and causing a strong distortion in the reconstruction.
Therefore we use the uniformity in the binned and normalized $R_{\rm rec}^{2}$ distribution, $P(R_{\rm rec}^{2})$, of the $^{83m}$Kr data.
Only $^{83m}$Kr events with $R_{\rm rec} < R_{\rm crit}$ are used in the evaluation, and $R_{\rm crit}$ is determined by
\begin{equation}
    P|_{R_{\rm rec}^2 = R_{\rm crit}^2} = 0.2\cdot P|_{R_{\rm rec}^2=0},
\end{equation}
which corresponds to the {\color{black} bin where the normalized distribution falls to 20\% of its central value}.

The relative standard deviation (RSD) in $P(R_{\rm rec}^{2})$ with $R_{\rm rec} < R_{\rm crit}$ is calculated after binning. 
Specifically, we set the binning as
\begin{equation}\label{equ:bin}
    R_{\rm rec}^{2}=1200\cdot n ~{\rm{mm^2}},~0\le {n} \le 100.
\end{equation}
The RSD is calculated as follows, 
\begin{equation}
\begin{split}
    {\rm RSD} &=\frac{\sqrt{\overline{P(n)^2} - [\overline{P(n)}]^2}}{\overline{P(n)}},~~{\rm~where} \\
     \overline{P(n)}&=\frac{\sum\limits_{n=0}^{n_{\rm max}}P(n)}{n_{\rm max}} ,~~\overline{P(n)^2}=\frac{\sum\limits_{n=0}^{n_{\rm max}}P(n)^2}{n_{\rm max}}
\end{split}
\end{equation}
and $n_{\rm max}$ corresponds to $R_{\rm crit}$.
For different group parameters, the center of the TPC share similar $R_{\rm rec}^2$ distribution where the reflection has little influence. RSD quantifies the uniformity extended to the edge but not influenced by a small amount of events reconstructed extremely outward. In general, the smaller the RSD is, the more uniform the $R_{\rm rec}^2$ distribution. 
 In Tab.~\ref{tab:cw}, the RSD with $c=1$, and $(d,w_{e})=(0.20,0.015)$ leads to the best performance. Then, the parameter $c$ in Eq.~\ref{equ:rho} is tuned with $d$ and $w_{e}$ slightly modified.

\begin{table}[htb!]
\centering
\caption{Values of RSD with $c=1$ and different $w_{e}$ and $d$.}
\label{tab:cw}
\begin{tabular}{l| l l l l l l l}
\hline\hline
\diagbox{$d$}{RSD}{$w_{e}$} & 0.010 & 0.015 & 0.020 \\
\hline
0.15 &0.146	&0.160 & 0.152\\
0.20	&0.123 & 0.115 & 0.143 \\
0.25 &0.126 &0.127& 0.134 \\

\hline \hline
\end{tabular}
\end{table}

RSD is minimized when $c=1.0$, $d=0.20$, $w_{e}=0.015$. If we change the calculation criteria of RSD such as the binning of $R_{\rm rec}^2$ and choice of $n_{\rm max}$, the best parameters will be slightly different, and cause around 1~mm difference in the reconstructed positions, which is minor compared to other uncertainties (see Sec.~\ref{sec:comp}). The final $\{\rho_{I},w_{I}\}$ for different groups is shown in Tab.~\ref{tab:rhoI}.

\begin{table}[htb!]
\centering
\caption{Tuned group parameters.}
\label{tab:rhoI}
\begin{tabular}{l l l l l l l l}
\hline\hline
$I$ & 1 & 2 & 3 & 4 & 5 & 6 & 7\\
\hline
$\rho_{I}$	&0.0059 & 0.0532& 0.0946& 0.0580& 0.0958&0.1703 &0.2\\
$w_{I}$	&0 & 0 & 0 & 0 & 0.015 &0.015&0.015\\
\hline \hline
\end{tabular}
\end{table}

\begin{figure}[!htb]
\centering
\includegraphics[width = 3.3in]{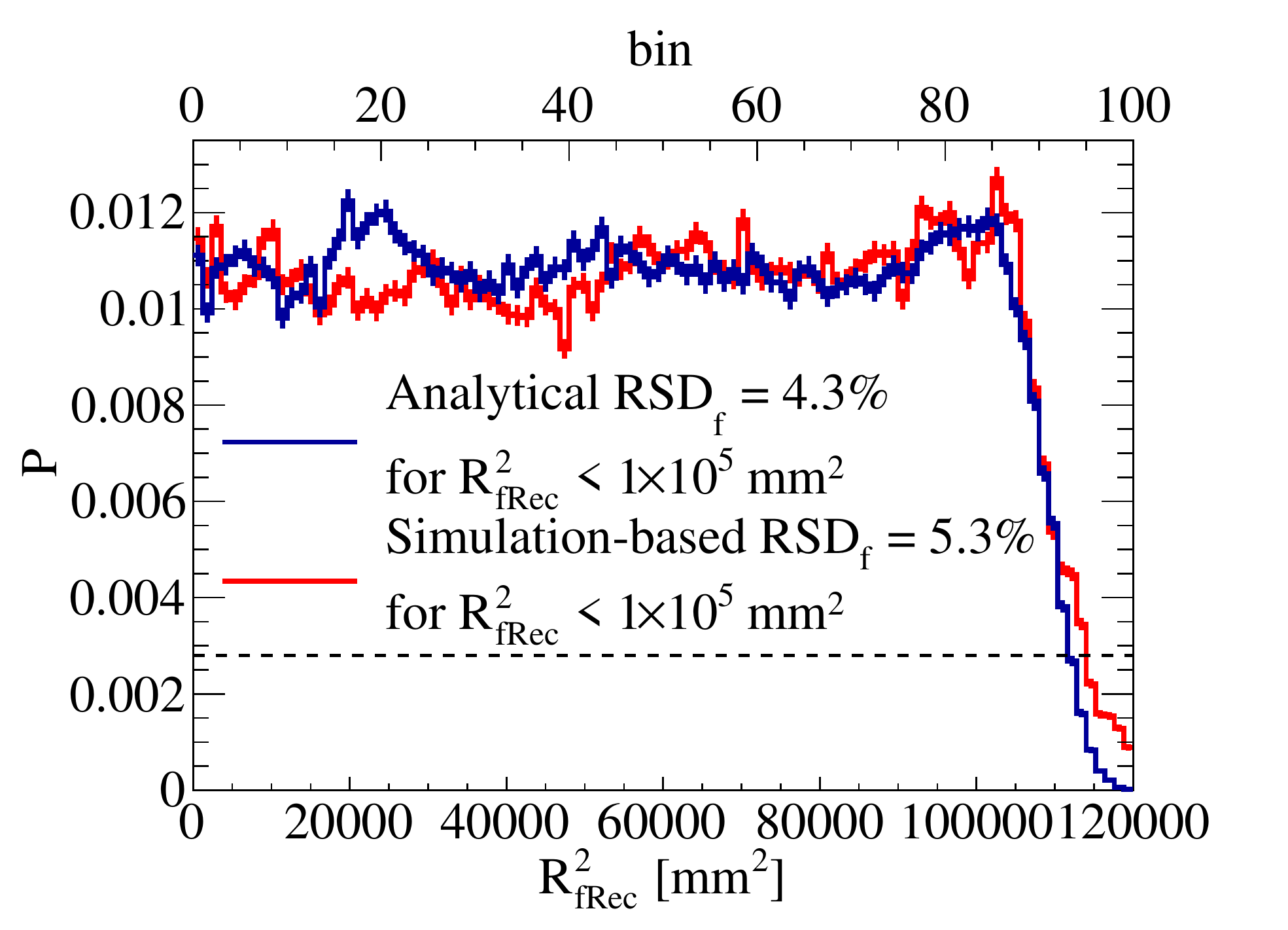}
\caption{$R_{\rm fRec}^{2}$ distribution of $^{83m}$Kr along $R_{\rm fRec}^2$ using the analytical (blue) and simulation-based (red) algorithms. The dashed black line is $0.2P|_{R_{\rm rec}^{2}=0}$ used in the analytical algorithm. The peaks along $R_{\rm fRec}^2$ in the analytical algorithm corresponding to the PMT centers (see Fig.~\ref{fig:2dPAF}). }
\label{fig:pafF}
\end{figure}

A modification is required to place surface events at the physical wall as discussed in \cite{surface}. A stretching factor of 1.07 is further applied on $(x_{\rm rec},y_{\rm rec})$. This might bring potential distortion in the reconstruction but is not significant compared to the local uncertainties (Sec.~\ref{sec:comp}).
 The $R_{\rm fRec}^2$ distributions of $^{83m}$Kr data with four inhibited PMTs is shown as the blue line in Fig.~\ref{fig:pafF}, where the `fRec' subscript stands for the final stretched reconstructed positions. The RSD$_{f}$ in $R_{\rm fRec}^2$ achieves 4.3\% as calculated by  Eq.~\ref{equ:bin} after replacing $R_{\rm rec}$ with $R_{\rm fRec}$ and $n_{\rm max}$ with $n_{f,{\rm max}}$ corresponding to $R_{\rm fRec}^{2} = 1\times10^{5}$~mm$^{2}$. The peaks along the $R_{\rm fRec}^2$ distribution are caused by reconstructed events gathering at the center of the PMTs, which is a minor problem as discussed in Sec.~\ref{sec:comp} (Fig.~\ref{fig:krcomp}).

\section{Simulation-based PAF}\label{sec:simu}
A Monte Carlo simulation, including event generator, light propagation, and signal reconstruction, is developed using the GEANT4 package for the PandaX-II detector~\cite{prd}.  
The event generator is a point source in the gas-phase immediately above the liquid and the photon numbers of the events follow Gaussian distribution. 
We fix the mean of the distribution as 10000~PE and sigma as 5000~PE to cover the region of interest.
The geometry follows the description in~\cite{prd}, and many parameters can be tuned for the light propagation, including the absorption length of photons in xenon, Rayleigh scattering length in xenon, reflection of the PTFE wall, etc. 
The output of this light simulation is the number of photons detected by each PMT for each event.

\begin{figure}[!htb]
\centering
\includegraphics[width = 3.5in]{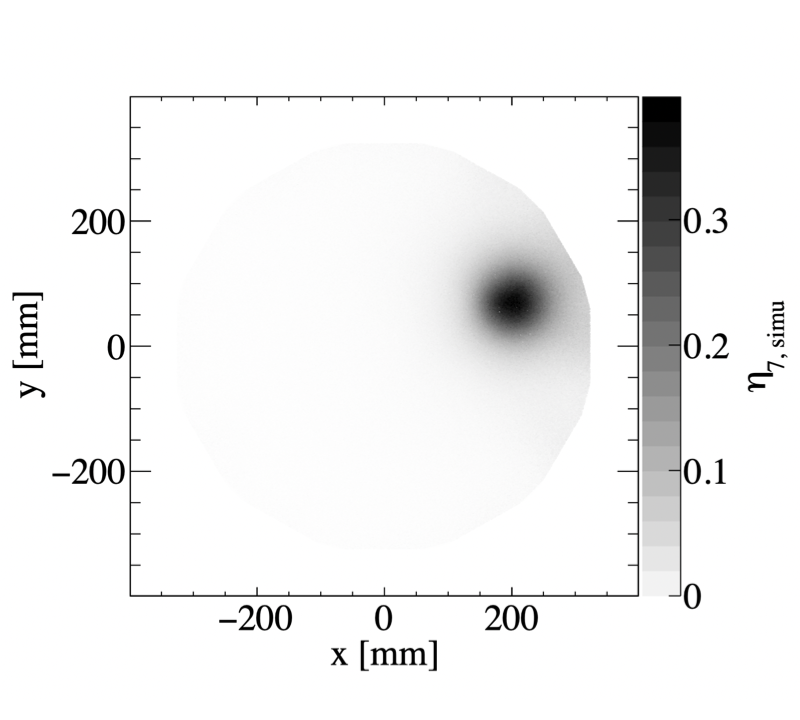}
\caption{An example of simulation-based PAF in the PandaX-II detector with half-million events. The maximum of $\eta_{7,{\rm simu}}$ is at the center of PMT$_{7}$ in Fig.~\ref{fig:pmtArray}.}
\label{fig:PAFex}
\end{figure}

To generate smooth PAFs, several hundred thousand events should be simulated uniformly at the liquid-gas interface, and the PAF$_i$ is evaluated using a formula similar to Eq.~\ref{equ:data},
\begin{equation}\label{equ:nn}
{\eta}_{i,\rm{simu}}(x,y)= \overline{ \frac{S2_{i,\rm{simu}}}{S2_{\rm{top,simu}}}}(x,y),
\end{equation} 
where $S2_{\rm{top,simu}}=\sum_{i}S2_{i,\rm{simu}}$ and the summation only includes top PMTs. The overline averages out the statistical fluctuation of the events in the same $x$-$y$ bin. An example of the simulation-based PAF$_7$ is shown in Fig.~\ref{fig:PAFex}. As the diameter of the PandaX-II detector is 658~mm (Fig.~\ref{fig:pmtArray}), for a 5~mm wide square bin, half million events are required for about 25 events in each bin.

It takes several hours to generate  simulation-based PAFs for one set configuration and reconstruct positions according to Eq.~\ref{equ:ml}. If there are enough computation resources, we can simulate all the possible configurations, and select the one with the best quality.

However, as mentioned before, computation requirement grows exponentially with the number of light propagation parameters. 
To find the most effective parameters within tolerable time, we focus on events at specific positions, the PMT centers. 
We select the $^{83m}$Kr events with the maximum of the normalized $\{S2_{i}\}$ hit pattern detected by the top PMTs larger than a preset value for each PMT$_{i}$.
The equivalent cut is $\eta_{i,\rm{data}}>t_{i}$, where $t_{i}$ is the preset value for PMT$_i$ and typically ranges from 0.3 to 0.5. 
The surviving events have the highest light fraction in the PMT$_{i}$. 
We assume that these events are under the center of the PMT$_i$ which generate an averaged data template at $(x_{\rm rec},y_{\rm rec})=(X_{i},Y_{i})$ without any reconstruction. 
With the simplification, we only need to simulate 100 events at each $(X_{i},Y_{i})$. Considering 51 PMTs are used (4 turned off), around five thousand events are enough for each configuration in the PandaX-II case. 

After many trials, we find that the light emission point of $S2$ in the gas phase, as a function of horizontal position, is an effective parameter to make the simulation agree with the data templates.
The vertical position, $z$, directly changes the angular coverage in the PMTs, which is more effective than the tuning of the reflectivity of the PTFE, the Rayleigh scattering length and absorption length. 
However, $z$ does not represent the real average positions of the proportional scintillation in the gas phase, as discussed in more detail at the end of this section.

The optimization of $z$ at a specific $(X_i, Y_i)$ is done with $\chi^{2}$ minimization,
\begin{equation}\label{equ:chi2}
\chi^{2}(z,X_i,Y_i)=\sum_{j}\frac{[{\eta}_{j,\rm{simu}}(X_i,Y_i|z)-{\eta}_{j,\rm{data}}(X_i,Y_i)]^2}{\sigma_{j,\rm{data}}(X_i,Y_i)^{2}},
\end{equation}
where $j$ is the PMT index.
Because the $(X_i, Y_i)$ position is fixed, $\eta_{j,\rm{simu}}$ is tuned as a function of $z$. The uncertainty $\sigma_{j,\rm{data}}$ is evaluated as
\begin{equation}
    \sigma_{j,\rm{data}}(X_i,Y_i)=\sqrt{\overline{(S2_{j}/S2_{\rm top})^2} - \left(\overline{S2_{j}/S2_{\rm top}}\right)^2},
\end{equation}
where the overline represents the average over the events selected at the center of PMT$_i$. 

\begin{figure}[!htb]
\centering
\includegraphics[width = 3.5in]{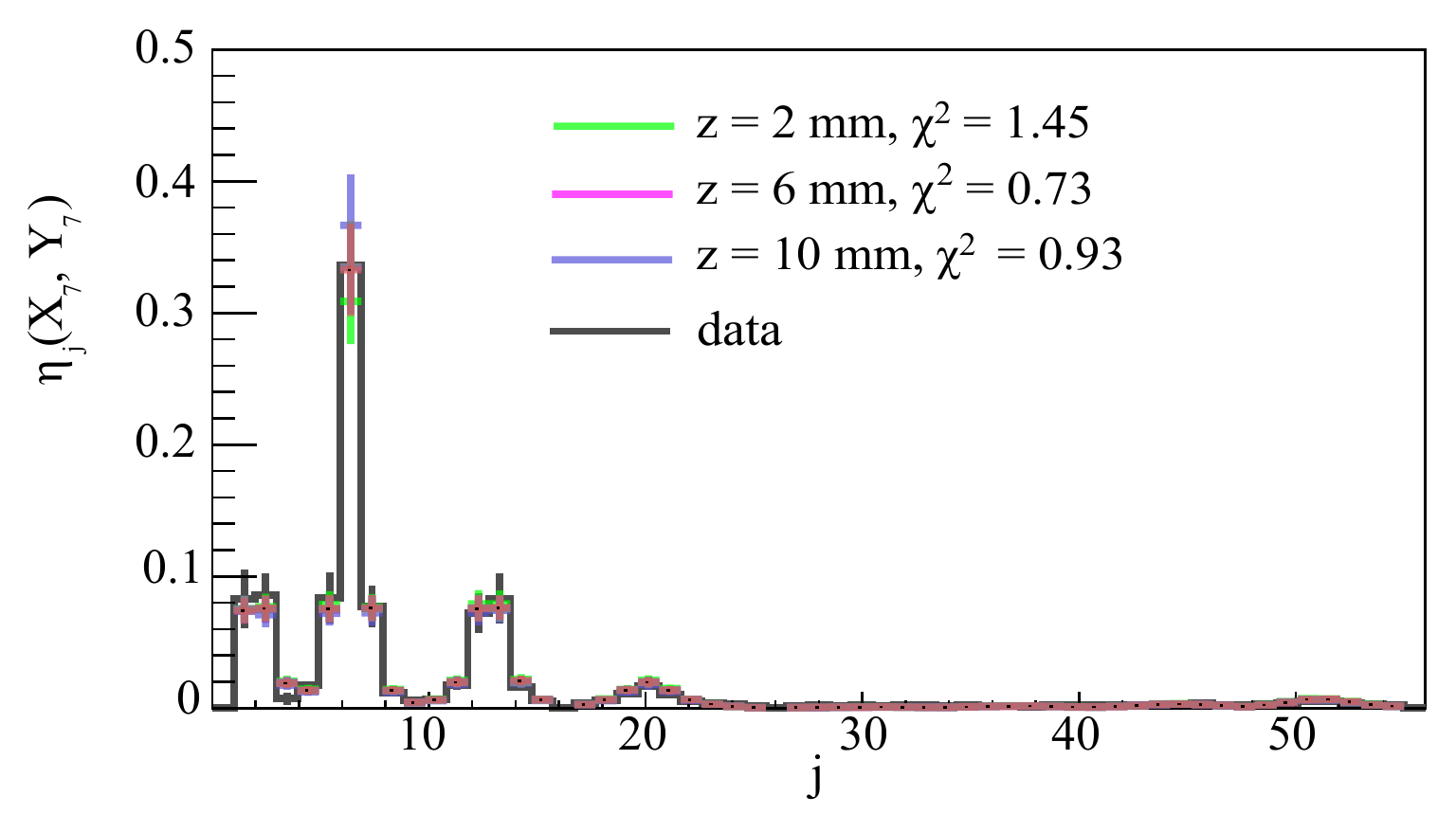}
\caption{An example of the data and simulation $\eta_j$ comparison at the center of PMT$_{7}$, $(X_7,Y_7)=(204,71)$~mm. The simulation templates are generated with different $z$ heights for the light emission..}
\label{fig:tuneNN}
\end{figure}

An example of tuning at the center of PMT$_7$ is plotted in Fig.~\ref{fig:tuneNN}. 
The $^{83m}$Kr events are selected by constraining $\eta_{7,\rm{data}}>0.4$. The $\sigma_{j,\rm{data}}(X_7,Y_7)$ is shown as error bars in the black histogram.
 In this example, $z=6$~mm is the one with the minimum $\chi^2$, and the liquid-gas interface corresponds to $z=0$. Moreover, compared to $z$, the reflectivity, $r$, is of a higher order as in Tab.~\ref{tab:zReflectivity}. In this GEANT4 simulation example, the reflectivity of the PTFE surface is modelled using a `ground' PTFE surface, `dielectric\_metal' interface and `SigmaAlpha=0.1' under the `unified' mode. The meanings of the keys are defined in the literature~\cite{g4_1,g4_2}. Because $r$ is a higher order effect, we use our measurement of the PTFE material reflectivity, 0.95, in the simulation~\cite{reflectivity}.

\begin{table}[htb!]
\centering
\caption{The $\chi^2$ dependence on the reflectivity $r$ with $z=6$~mm at $(X_7,Y_7)$.}
\label{tab:zReflectivity}
\begin{tabular}{ l l l l l}
\hline\hline
$r$ & 0.25 &0.50 &0.75 &1.00\\
 $\chi^2$& 0.75 & 0.74 & 0.73 & 0.78\\
\hline \hline
\end{tabular}
\end{table}

\begin{figure}[!htb]
\centering
\includegraphics[width = 3.5in]{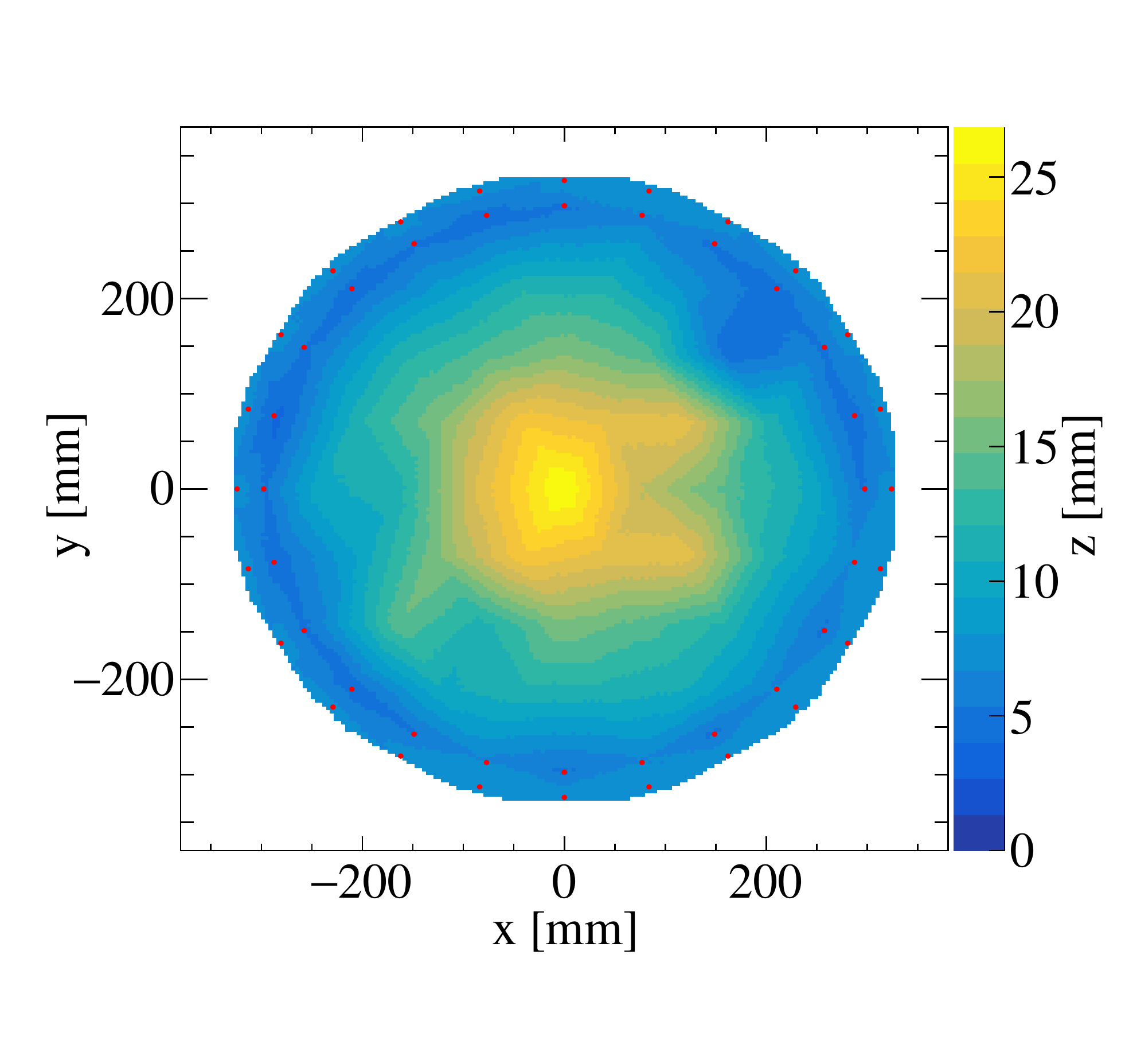}
\caption{The optimized surface of $z(x,y)$ (the effective vertical position of  $S2$ emissions) for the PandaX-II detector with 2D linear interpolation applied. The red circles are $\{(x_{k}, y_{k})\}$ which used to optimize $z(x,y)$ at the edge.}
\label{fig:zNN}
\end{figure}

However, the optimization of $z$ at the PMT centers is not enough for the area close to the PTFE surface at large $R$. A iterative tuning is performed. Starting with $z$ at the liquid-gas interface, we simulate half-million events horizontally uniformly distributed in the gas phase to generate PAFs, and they are used to reconstruct the calibration data. 
Following this, a set of target positions which are marked as red circles in Fig.~\ref{fig:zNN}, $\{(x_{k}, y_{k})\}$, is chosen to represent the local behaviors. 
For each $(x_{k}, y_{k})$, 100 $^{83m}$Kr events reconstructed closest to it are averaged to represent the reconstructed data. 
Then, the simulation at the same $(x_{k}, y_{k})$ position with different $z$ are done. 
The new $z$ with the best agreement to the data is updated for each target position. Two or three iterations are enough to find the optimized $z(x_{k},y_{k})$. After optimizing the $z$ at different points, we use 2D linear interpolation to generate the mapping, $z(x,y)$, as in Fig.~\ref{fig:zNN}. 
We generate the final half million events in the gas phase with the optimized surface of $z(x,y)$.

The shape of $z(x,y)$ reveals some physical effects on the PAFs.
The reflection of the PTFE surface leads to a small increase in $z$ at the border.
The center region has a highest $z$ which may be caused by larger $E_{\rm extraction}$ due to the deformation of the electrodes. Nevertheless, the deformation of the grid wires on the gate electrode should be sub-milimeter as suggested in another simulation~\cite{reflectivity}. Moreover the distance between the gate and anode is 11~mm which is smaller than the $z$-parameter at the center. Therefore, $z$ are not the real positions of $S2s$. A more reasonable explanation is that the non-uniform $E_{\rm extraction}$ results in different $S2$ responses horizontally even for mono-energetic gammas~\cite{main2,recentMain}, and the PAFs are different correspondingly.  

A stretching factor, 1.06, determined by the PTFE surface events in \cite{surface}, is applied to the original reconstructed positions. The factor is slightly different from the analytical algorithm because the two present different radial bias as in Fig.~\ref{fig:krcompb}. The $R_{\rm fRec}^{2}$ distribution of the $^{83m}$Kr data with four PMTs turned off are shown in Fig.~\ref{fig:pafF}. A 5.3\% RSD$_{\rm fRec}$ in the $R_{\rm fRec}^2$ distribution of the $^{83m}$Kr data is reached in $R_{\rm fRec}^2<1\times10^{5}$~mm$^{2}$.

\section{Comparison}\label{sec:comp}
In this section, we compare the position reconstruction algorithms regarding uniformity, robustness and uncertaitnties. 
Before the comparison, we first estimate the best performance of the ML (Eq.~\ref{equ:ml} in Sec.~\ref{sec:paf}) by using simulation-based PAFs to reconstruct the corresponding simulation data. The uniformity and robustness are evaluated with $^{83m}$Kr events. 
 We calculate the uncertainties in the center area by the radioactivities on the gate electrode. The uncertainties at the border are estimated by the PTFE surface events as in \cite{surface}.

\begin{figure}[!htb]
\centering
\subfloat[\label{limitDeltaRabs}]{
\label{fig:drsimua}
    \includegraphics[width=3.5in]{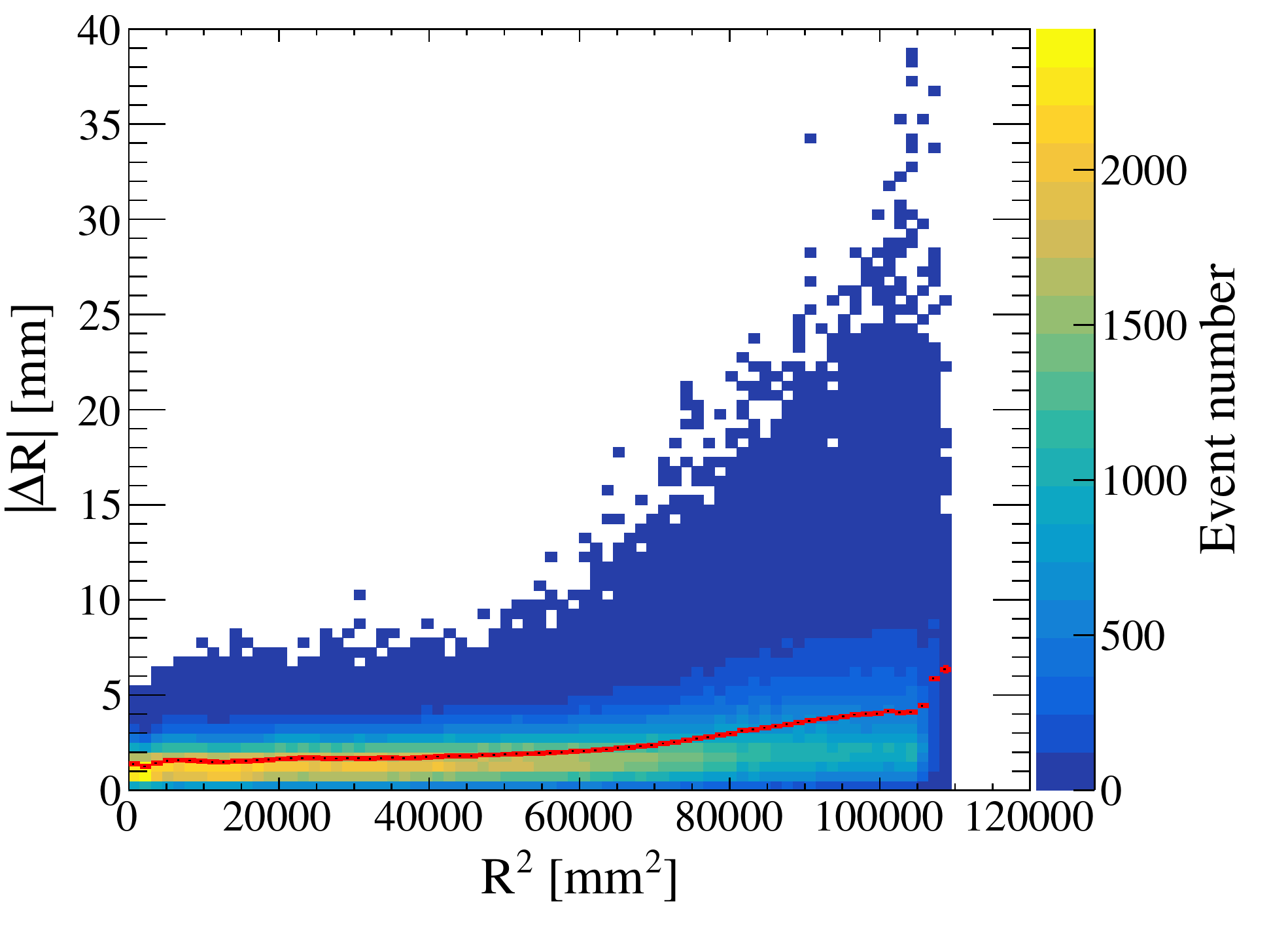}}
  
  \subfloat[\label{limitDeltaR}]{
  \label{fig:drsimub}
    \label{fig:limitdeltaR}
    \includegraphics[width=3.5in]{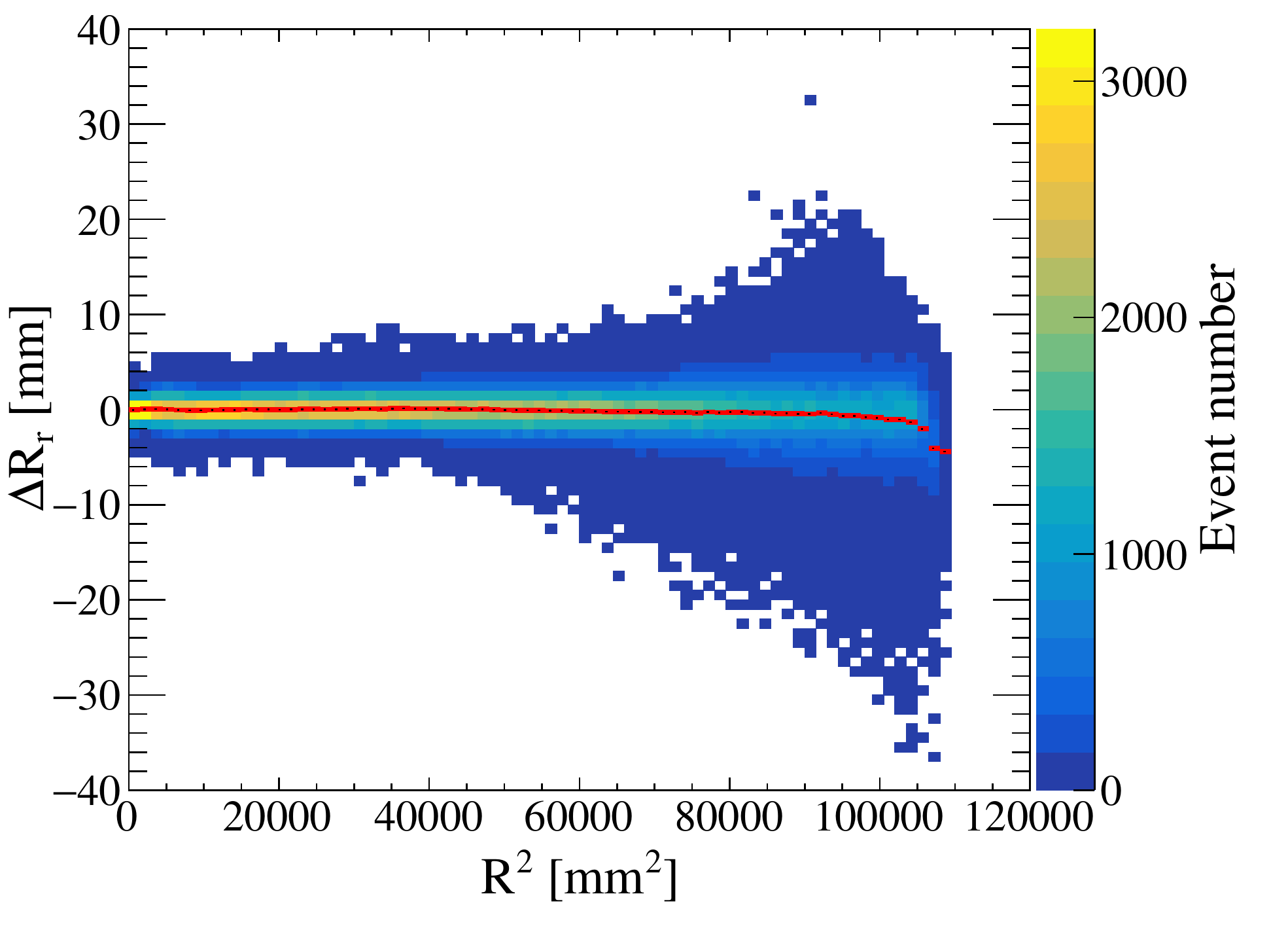}}
\caption{
The deviation between ML reconstructed (Eq.~\ref{equ:ml}) and true positions using simulated data (mean $S2$ at 10000~PE with a 5000~PE sigma) with corresponding simulated PAF,  
(a) Total deviation, $|\Delta R|$, vs $R^{2}$  (b) Radial deviation, $\Delta R_{r}$, vs $R^{2}$. The red line represents the mean deviation, and therefore serve as an estimate of the intrinsic uncertainty of the ML fit. }
\label{fig:drsimu}
\end{figure}

As the simulation data have known positions, the uncertainties can be directly evaluated by the difference between the reconstructed position $(x_{\rm rec},y_{\rm rec})$ and the origin ($x_{\rm true},y_{\rm true}$) as $|\Delta{R}|=\sqrt{(x_{\rm rec}-x_{\rm true})^2+(y_{\rm rec}-y_{\rm true})^2}$. The PAF is constructed by Eq.~\ref{equ:nn}, and the position is reconstructed by ML defined in Eq.~\ref{equ:ml}. The total deviation, $\Delta{R}$, is plotted as functions of $R^{2}$ in Fig.~\ref{fig:drsimua}, and $\overline{\Delta{R}}$ is 2.2~mm including all simulated data. The radial deviation, $\Delta{R}_r=\sqrt{x_{\rm rec}^2+y_{\rm rec}^2}-\sqrt{x_{\rm true}^2+y_{\rm true}^2}$, in Fig.~\ref{fig:limitdeltaR} shows a systematic inward deviation at large $R$. The best performance is limited by the hardware setup, including the horizontal distances among PMT centers and the vertical distance from the liquid-gas interface to the top-array PMTs. The border area performs worse because of a less angular coverage.

\begin{figure}[!htb]
\centering
\subfloat[\label{fig:deltaRabs}]{
\label{fig:krcompa}
    \includegraphics[width=3.5in]{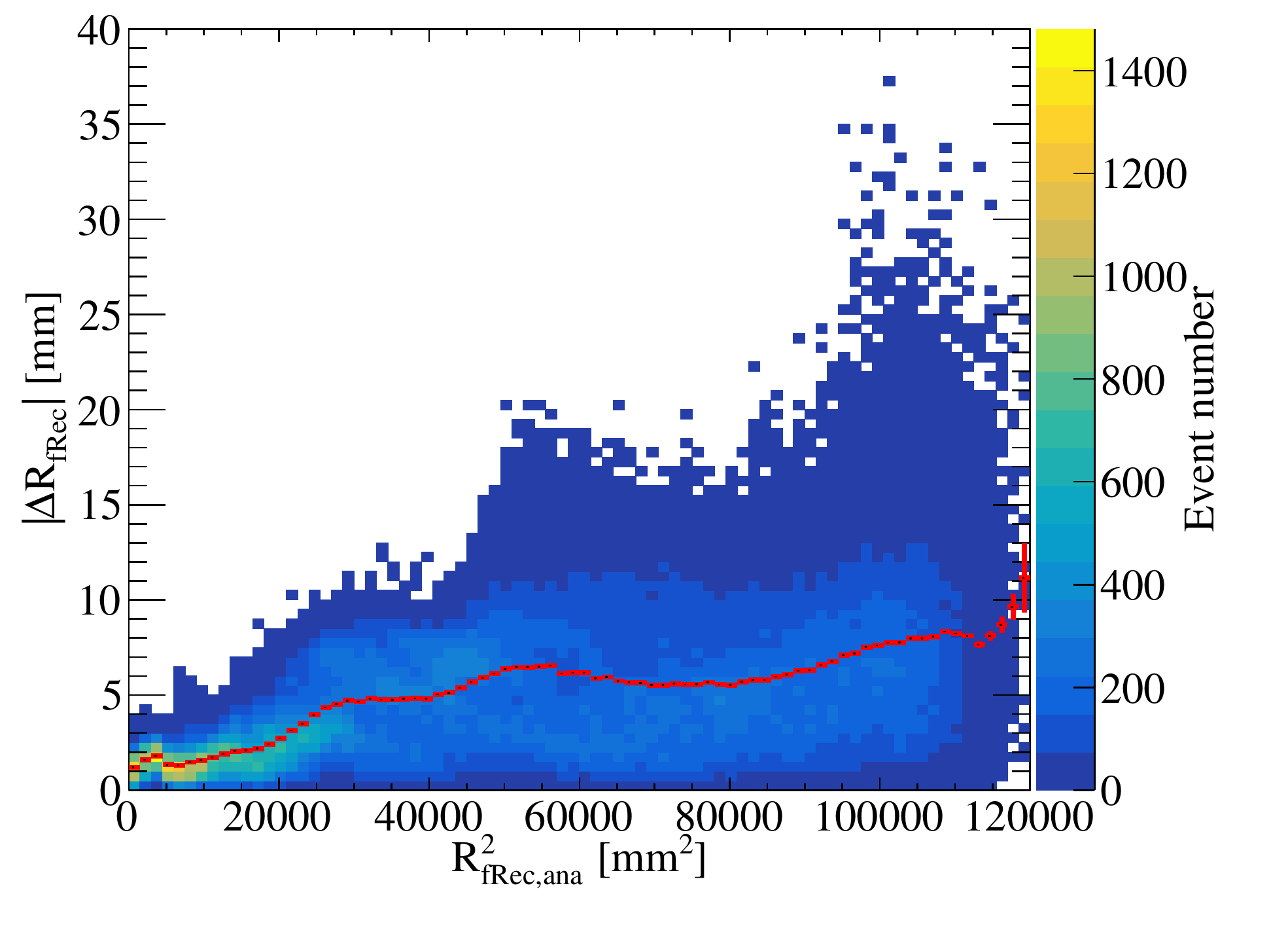}}
  
  \subfloat[\label{fig:deltar}]{
  \label{fig:krcompb}
    \includegraphics[width=3.5in]{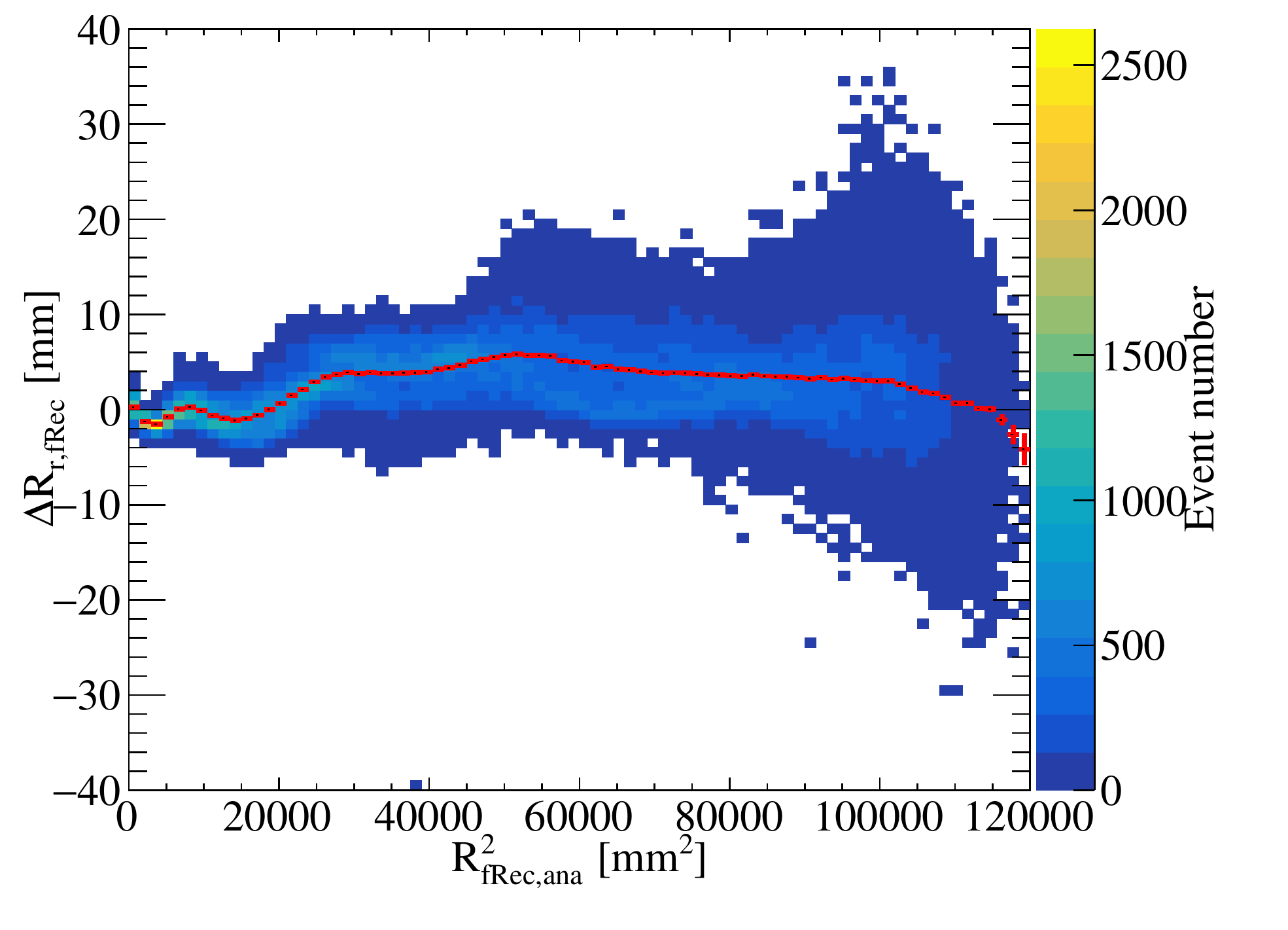}}
\caption{ The total deviation between the reconstructed positions obtained by the two algorithms for the $^{83m}$Kr events. (a) Total deviation, $|\Delta{R}_{\rm fRec}|$, vs $R_{\rm fRec,ana}^{2}$. (b) The radial deviation, $\Delta R_{r,\rm fRec}$, $R_{\rm fRec,simu} - R_{\rm fRec,ana}$, vs $R_{\rm fRec,ana}^{2}$. In both figures, the x-axis, $R_{\rm fRec,ana}^{2}$, is calculated with the analytical algorithm. The red line again represents the mean deviation.}
\label{fig:krcomp}
\end{figure}

The total deviation between the two algorithms in the reconstructed real $^{83m}$Kr events, 
\begin{equation}
    |\Delta R_{\rm fRec}| =\sqrt{(x_{\rm fRec,ana}-x_{\rm fRec, simu})^2 +(y_{\rm fRec, ana}-y_{\rm fRec,simu})^2},
\end{equation}
  is shown in Fig.~\ref{fig:krcompa}. The average total deviation, $\overline{|\Delta R_{\rm fRec}|}$, throughout the plane is $(5.2\pm3.6)$~mm, which reflects the propagated errors of the local uncertainties in the two algorithms and the distortion due to the surface events stretching. Similar to Fig.~\ref{fig:drsimu}, the outer part is worse. Moreover, the clustering of events under the PMT centers in the analytical algorithm is reflected in the fluctuation along  $R_{\rm fRec,ana}^{2}$, which is minor compared to the absolute deviation. The radial deviation is calculated as 
  \begin{equation}
      \Delta{R}_{r,\rm fRec}=\sqrt{x_{\rm fRec,simu}^2+y_{\rm fRec,simu}^2}-\sqrt{x_{\rm fRec,ana}^2+y_{\rm fRec,ana}^2}.
  \end{equation}
  The $\Delta{R}_{r,\rm fRec}$ distribution along $R_{\rm fRec,ana}^2$ in Fig.~\ref{fig:krcompb} reveals that the simulation-based algorithm reconstructed events slightly more outward in $(30000,11000)$~mm$^2$ with a peak around $50000$~mm$^2$.

\begin{figure}[!htb]
  \centering
  \subfloat[\label{fig:2dPAF}]{
    \includegraphics[width=4.3in]{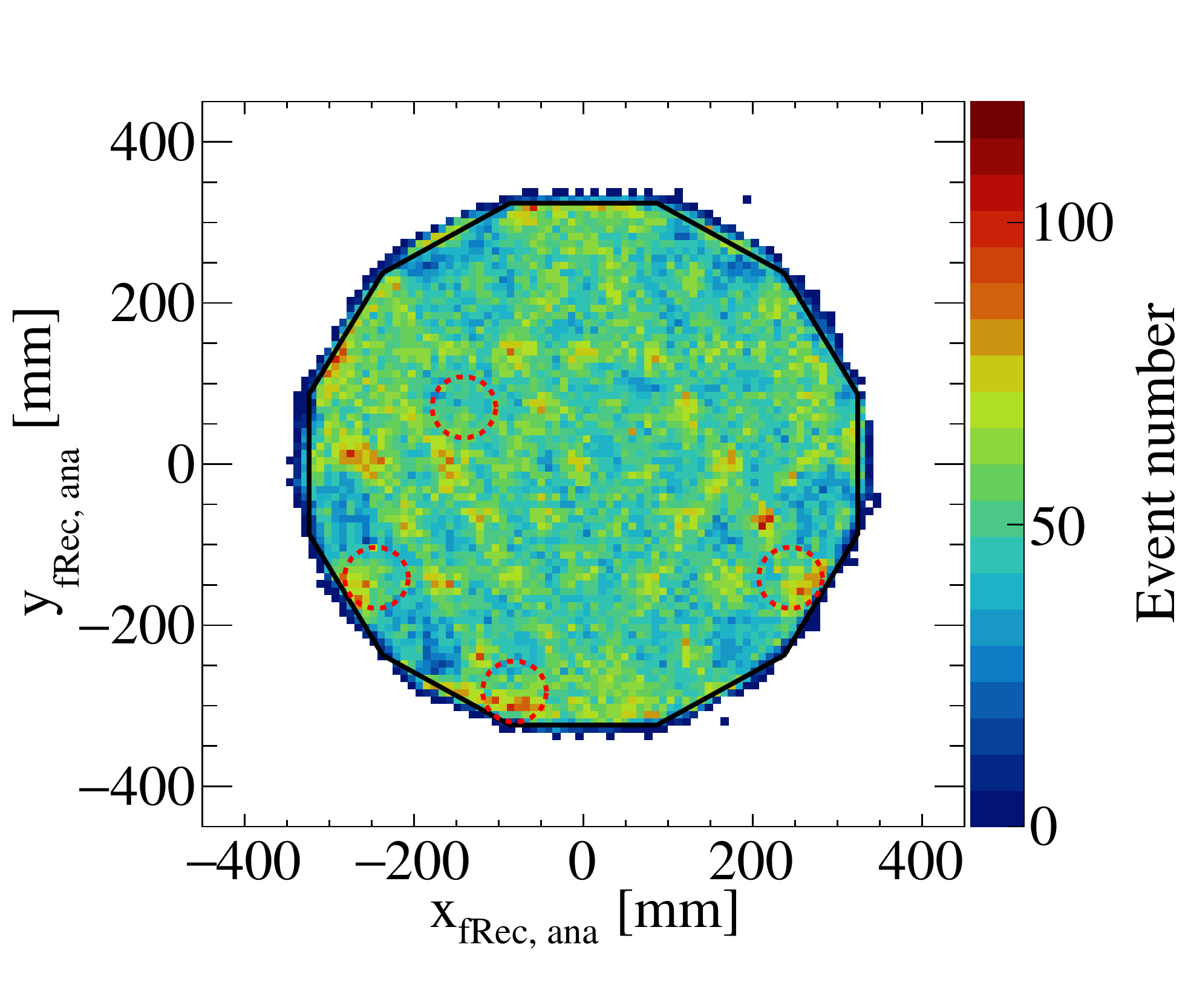}}
  
  \subfloat[\label{fig:2dnn}]{
    \includegraphics[width=4.3in]{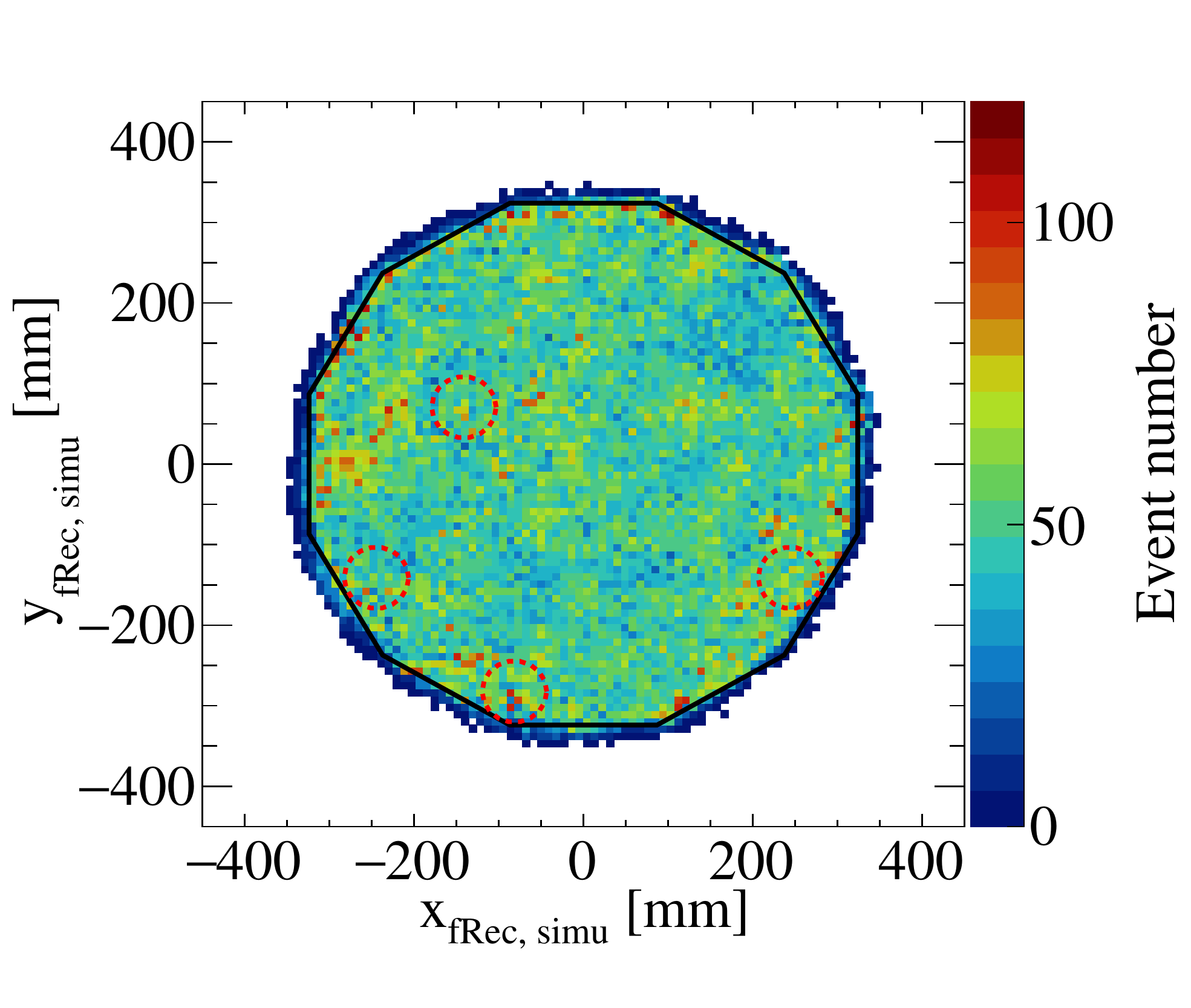}}
  
  \caption{ $x_{\rm fRec}-y_{\rm fRec}$ $^{83m}$Kr event distribution using the (a) analytical and (b) simulation-based algorithms. The four red circles mark the inhibited PMTs.}
  \label{fig:2dComp} 

\end{figure}

The uniformity comparison in the $R_{\rm fRec}$ distributions of the $^{83m}$Kr events is shown in Fig.~\ref{fig:pafF}, where the analytical algorithm wins over the simulation-based slightly. In the $x_{\rm fRec}$-$y_{\rm fRec}$ distribution (Fig.~\ref{fig:2dComp}), the average of the number of $^{83m}$Kr events over the bins within the detector boundary (black line) is $49.4\pm13.4$ ($48.9\pm 14.1$) for the analytical (simulation-based) algorithm. The standard deviations of the event numbers are taken as uncertainties in the average. The 2D uniformity of $^{83m}$Kr is consistent with the $R_{\rm fRec}$ distribution. 

The robustness in the simulation-based algorithm is slightly better than the analytical one. The average of the $^{83m}$Kr event number calculated similarly for the third quadrant in Fig.~\ref{fig:2dComp} where two close PMTs are turned off is $48.8\pm14.3$ ($48.4\pm 13.9$) for the analytical (simulation-based) algorithm. The change of standard deviations suggests that the simulation-based algorithm is more stable when handling inhibited PMTs.

\begin{figure}[!htb]
\centering
\includegraphics[width = 3.5in]{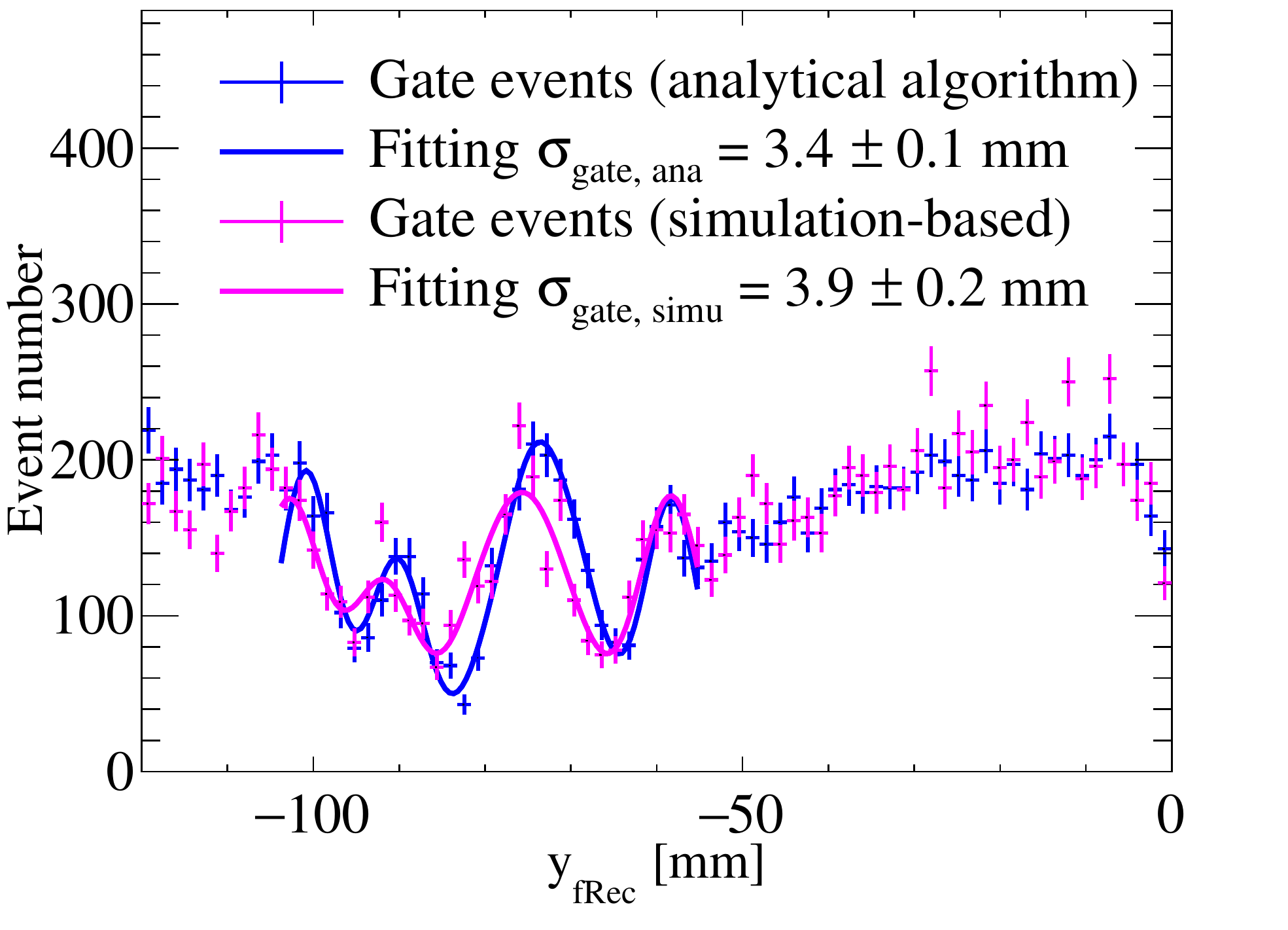}
\caption{The gate events $y_{\rm fRec}$ distribution projected along the grid wires with the analytical (blue) and  simulation-based (magenta) algorithms. The solid lines are the corresponding fittings to the $y_{\rm fRec}$ distribution by six Gaussian functions with the same width.}
\label{fig:gate}
\end{figure}

The $y_{\rm fRec}$ distributions of the gate-electrode events with $S2\in(1000,20000)$~PE are used to study the uncertainties in the center of the detector. {\color{black}These events are likely to come from radioactive isotopes attached to the grid wires. Because the electric field close to the wires is quite different from the bulk sensitive region, we lack enough information to identify the dominant isotopes.} The gate events can be selected with the characteristic drifting time ($2$-$4$~$\mu$s). In Fig.~\ref{fig:gate}, we also apply a spatial cut, $R_{\rm fRec}^2 < 72000$~mm$^2$, to suppress backgrounds from the surface and outside the detector. This fiducial radius cut is the same as the WIMP search in PandaX-II~\cite{main2}. 

The grid wires are parallel to the $x$-axis and 5~mm apart along the $y$-axis. The gaps normally are not recognizable as shown by the events with $y_{\rm fRec}\in(-50,0)$~mm in Fig.~\ref{fig:gate}. For clarity, the peaks in the simulation-based algorithm in this region are caused by the binning of the 2D PAFs. Occasionally, due to some local defects, larger gaps can be seen. We identify the malfunctioning grid wires according to the troughs and fit the $y_{\rm fRec}$ distribution with Eq.~\ref{equ:gate1d}, where $p_0$ is constrained to the peak within $(-95,-85)$~mm, $\sigma_{\rm gate}$ represents the uncertainties in the position reconstruction, $d_0$ is the reconstructed gap between grid wires, and $N_1$ to $N_6$ are the fitted event numbers on the corresponding grid wires.
The three troughs in Fig.~\ref{fig:gate} at $-95$~mm, $-85$~mm and $-65$~mm correspond to the Gaussian functions skipped for the centers at $p_{0}-1d_{0}$, $p_{0}+1d_{0}$ and $p_{0}+5d_{0}$, respectively. At these sites, the grid wires may be sagging or have poor electrical connections with their holder in $-100$~$^{\circ}$C liquid xenon.

\begin{equation}\label{equ:gate1d}
\begin{split}
f = ~& {\rm{Gaus}}(y_{\rm fRec},p_0,\sigma_{\rm gate})\cdot N_1 +\\
&{\rm{Gaus}}(y_{\rm fRec},p_0-2d_0,\sigma_{\rm gate})\cdot N_2 +\\
&{\rm{Gaus}}(y_{\rm fRec},p_0+2d_0,\sigma_{\rm gate})\cdot N_3 + \\
&{\rm{Gaus}}(y_{\rm fRec},p_0+3d_0,\sigma_{\rm gate})\cdot N_4 +\\ &{\rm{Gaus}}(y_{\rm fRec},p_0+4d_0,\sigma_{\rm gate})\cdot N_5 +\\
&{\rm{Gaus}}(y_{\rm fRec},p_0+6d_0,\sigma_{\rm gate})\cdot N_6,
\end{split}
\end{equation}
and the Gaus is the Gaussian function,
\begin{equation}
\label{equ:gaus}
{\rm{Gaus}}(x,\mu,\sigma) = \frac{1}{\sigma\sqrt{2\pi}}\exp\left[-\frac{(x-\mu)^2}{2\sigma^{2}}\right].
\end{equation}
 
The best-fitting parameters are shown in Tab.~\ref{tab:gateFit}. As the diameter of the wires (100~$\mu m$) is much smaller than the $\sigma_{\rm gate}$, the dispersion represents the local uncertainties. $\overline{|\Delta R_{\rm fRec}|}$ in Fig.~\ref{fig:krcompa} ($5.2\pm3.6$~mm) is consistent with the propagated uncertainty $\sqrt{\sigma_{\rm gate, ana}^{2}+\sigma_{\rm gate, simu}^2} = 5.2$~mm. In principle, the event numbers, from $N_1$ to $N_6$, should be the same regardless of the reconstruction algorithms. $N_{1,2,6}$ for the two algorithms are consistent within the fitting errors. However, because we use three Gaussian functions for a single peak in $(-85,-65)$~mm to constrain $d_0$ better, which brings too many degrees of freedom, the differences in $N_{3,4,5}$ are larger.

\begin{table}[htb!]
\centering
\caption{The results of fitting Eq.~\ref{equ:gate1d} with the $y_{\rm fRec}$ of the gate events reconstructed with the two algorithms.}
\label{tab:gateFit}
\begin{tabular}{ l l l l l l l }
\hline\hline
                    & $\sigma_{\rm gate}$ [mm] & $d_{0}$ [mm] & $p_{0}$ [mm] & & &\\ 

          & $3.36\pm0.10$ & $5.33\pm0.04$ &$-90.2\pm0.2$  & & &\\
Analytical&$N_1$  & $N_2$  & $N_3$  & $N_4$  & $N_5$  & $N_6$    \\
&$135\pm7$  & $192 \pm 8$ & $57\pm 8$ & $165\pm 9$  & $100\pm10$ & $175 \pm 7$\\
\hline
& $\sigma_{\rm gate}$ [mm] & $d_{0}$ [mm] & $p_{0}$ [mm] & & &\\ 
    & $3.94\pm0.16$ & $5.58\pm0.07$ &$-91.7\pm0.3$ & & &\\
 Simulation-based & $N_1$  & $N_2$  & $N_3$  & $N_4$  & $N_5$  & $N_6$    \\
 &$119\pm7$  & $173 \pm 9$ & $84\pm 10$ & $125\pm 10$ & $60\pm10$ & $176 \pm 7$\\
\hline \hline
\end{tabular}
\end{table}



%

\begin{figure}[!htb]
\centering
\subfloat[]{
    \label{fig:surface_ex}
    \includegraphics[width=3.5in]{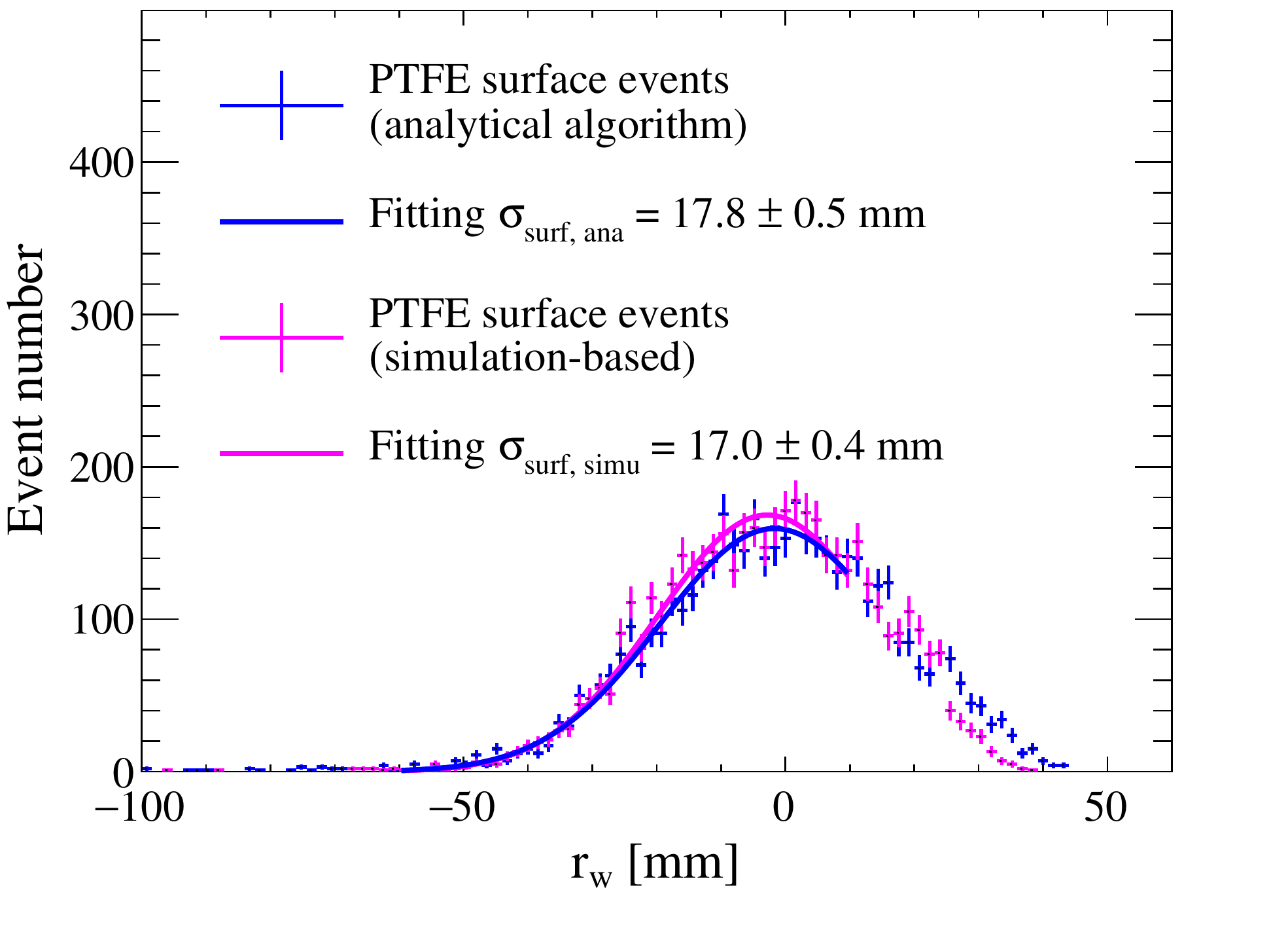}}
  
  \subfloat[]{
  \label{fig:surface_tot}
    \includegraphics[width=3.5in]{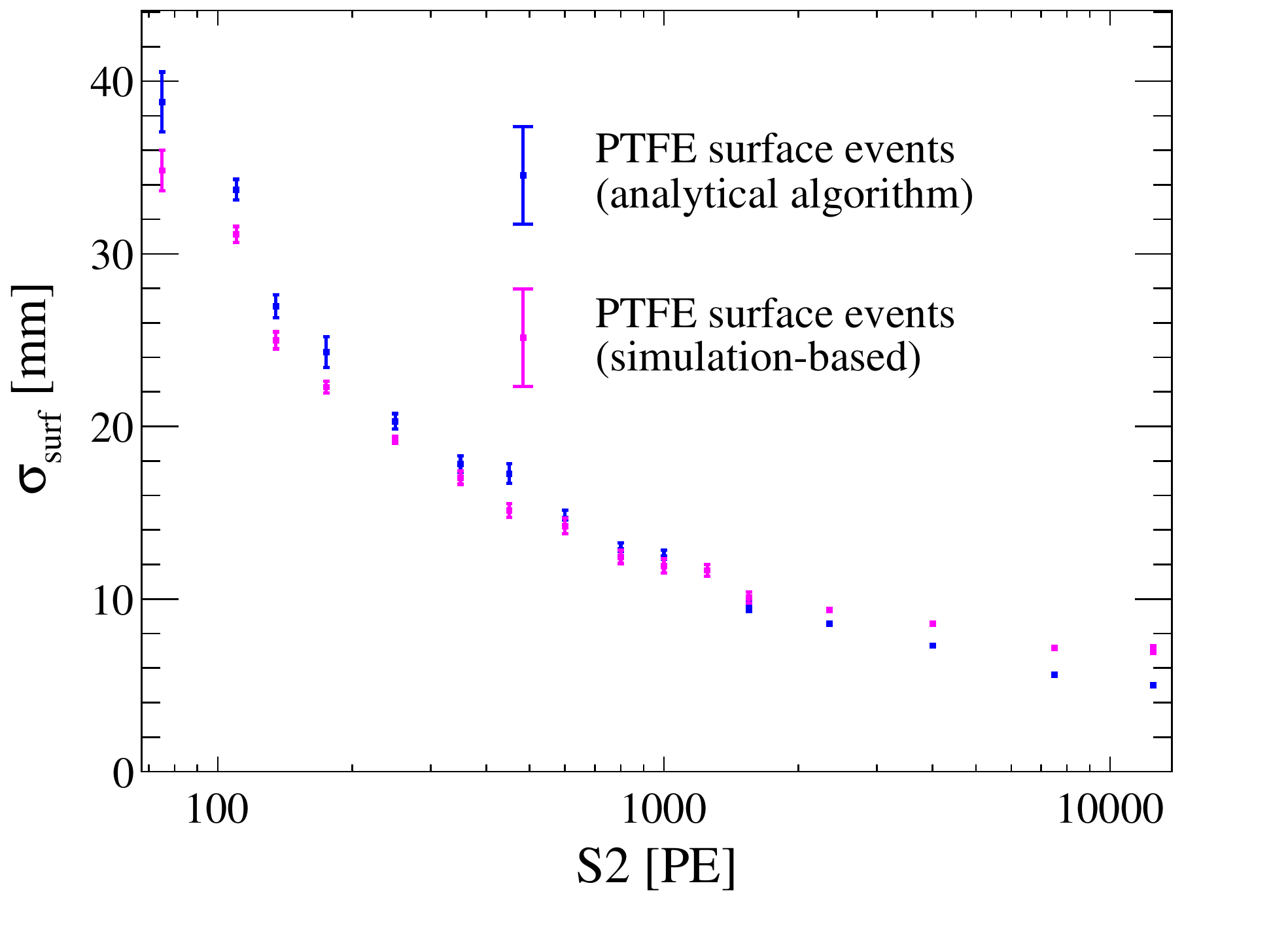}}
\caption{(a) The distribution of the PTFE surface events along the distance to the surface, $r_w$, with $S2\in (300,~400)$~PE. The solid lines are Gaussian functions fitted to the distributions with $r_w < 10$~mm (blue for analytical and magenta for simulation-based). (b) The fitted radial dispersion, $\sigma_{\rm surf}$, vs $S2$.}
\label{fig:surface}
\end{figure}

Furthermore, we use the PTFE surface events to study the upper bound of the uncertainties with different $S2s$~\cite{surface}. The surface events in the WIMP search signal window are dominated by $^{222}$Rn progenies plated out on the TPC inner surface~\cite{recentMain}.
An example of determining the uncertainties for the surface events with $S2\in (300,400)$~PE is presented in Fig.~\ref{fig:surface_ex}, where $r_w$ is the distance to the PTFE surface. We don't use $R_{\rm fRec}$ because the detector is not a cylinder, and a small inward shifting of the reconstructed surface with a larger drifting time is also corrected in $r_w$. The negative part of $r_w$ corresponds the sensitive region of the TPC, and is fitted to a Gaussian function as Eq.~\ref{equ:gaus} with $r_w <10$~mm. Because the width of the surface-event distribution is not the same on the two sides especially in the region $|r_{w}|> 30$~mm, we confine the Gaussian fitting to a small positive $r_w$ number. The width represents the uncertainties in the surface events. For $S2\in (300,400)$~PE, the analytical (simulation-based) algorithm has $\sigma_{\rm surf,~ana} = 17.8\pm 0.8$~mm ($\sigma_{\rm surf,~simu}=17.0\pm 0.4$~mm) with a center fitted at $-1.7$~mm ($-2.8$~mm). As shown in Fig.~\ref{fig:surface_tot}, the statistical uncertainties dominate when $S2$ approaches to smaller than 1000~PE in the PandaX-II detector.


In brief, the analytical algorithm has slightly smaller uncertainties with several thousand PEs in $S2s$ and better uniformity along $R_{\rm fRec}^2$ distribution, but slightly larger uncertainties with several hundred PEs in $S2s$ and less robustness when handling inhibited PMTs. In general, the two position reconstruction algorithms are comparable for the WIMP search purpose.


\section{Conclusion}
To reach a millimeter level resolution in the horizontal position reconstruction for a TPC mounted with 3-inch PMTs,
we develop two algorithms based on the previous works. 
In the analytical algorithm, we introduce two groups of parameters to extend the axially-symmetric PAF for non-negligible reflection. In the simulation-based one, we tune the light emission points in the gas phase as a function of $x$ and $y$. Both algorithms are trained with $^{83m}$Kr data. The reconstructed horizontal positions are stretched radially by a factor of $1.07$~($1.06$) in the analytical (simulation-based) algorithm to make the mean radius of the surface-event positions agree with the solid boundary.

Applying both algorithms to the PandaX-II detector, the uniformity of  the $R_{\rm fRec}^2$ distribution of $^{83m}$Kr reaches 4.3\% (5.3\%) in the analytical (simulation-based) algorithm within $R_{\rm fRec}^2< 1\times10^{5}$~mm$^{2}$, and the average difference in the reconstructed positions between the two algorithms is $5.2\pm3.6$~mm. Using the gate events, the uncertainties are $3.4$~mm ($3.9$~mm) for the analytical (simulation-based) algorithm when $S2s$ are of several thousand PEs. For $S2s$ with several hundred PEs, the uncertainties are several centimeters near the PTFE surface.

As two algorithms are comparable in uncertainties in the WIMP search $S2$ region, we decide to apply the simulation-based algorithm  in the final analysis of PandaX-II because of the robustness, and preserve the analytical algorithm as a crosscheck.

\section{Acknowledgement}
This project is supported in part by office of Science and Technology, Shanghai Municipal Government (grant No. 18JC1410200), a grant from the Ministry of Science and Technology of China (No. 2016YFA0400301), grants from National Science Foundation of China (Nos. 12005131, 11905128, 12090061, 11775141), and a grant from Sichuan Science and Technology Program (No.2020YFSY0057). We thank supports from Double First Class Plan of the Shanghai Jiao Tong University. We also thank the sponsorship from the Chinese Academy of Sciences Center for Excellence in Particle Physics (CCEPP), Hongwen Foundation in Hong Kong, and Tencent Foundation in China. Finally, we thank the CJPL administration and the Yalong River Hydropower Development Company Ltd. for indispensable logistical support and other help.

%





\end{document}